\documentclass[3p,twocolumn,times]{elsarticle}

\usepackage[utf8]{inputenc}
\usepackage[ngerman,english]{babel, varioref}

\usepackage{hyperref}
\usepackage{isotope} 
\usepackage{graphicx}
\usepackage{float}
\usepackage{booktabs}
\usepackage{amssymb}
\usepackage{amsthm}
\usepackage{amsmath}
\usepackage{multirow}

\usepackage[binary-units=true,abbreviations=true]{siunitx} 
\DeclareSIUnit\mwe{m w.e.}
\DeclareSIUnit\counts{counts}

\usepackage[figuresright]{rotating}

\begin{document}
\begin{frontmatter}

\title{The Dortmund Low Background Facility --- \\ Low-Background Gamma Ray Spectrometry with an Artificial Overburden}

\author[tudo]{Holger Gastrich}
\author[tudo]{Claus Gößling}
\author[tudo]{Reiner Klingenberg}
\author[tudo]{Kevin Kröninger}
\author[tudo]{Till Neddermann}
\author[tudo]{Christian Nitsch\corref{cor1}}  \ead{christian.nitsch@tu-dortmund.de}
\author[tudo]{Thomas Quante}
\author[tudd]{Kai Zuber}

\cortext[cor1]{Corresponding author. tel +49 231 755 8509; fax +49 231 755 3688}

\address[tudo]{Technische Universität Dortmund, Lehrstuhl für Experimentelle Physik IV \\ Otto-Hahn-Str. 4, 44221 Dortmund, Germany}
\address[tudd]{Technische Universität Dresden, Institut für Kern- und Teilchenphysik \\ Zellescher Weg 19, 01069 Dresden, Germany}

\pdfinfo{%
  /Title    (The Dortmund Low Background Facility --- Low-Background Gamma Ray Spectrometry with an Artificial Overburden)
  /Author   (C. Nitsch)
  /Creator  ()
  /Producer ()
  /Subject  ()
  /Keywords (Low-level background \sep HPGe detector \sep Gamma-ray spectrometry \sep Background reduction)
}

\begin{abstract}
High-purity germanium (HPGe) detectors used for low-background gamma ray spectrometry are usually operated under either a fairly low overburden of the order of one meter of water equivalent (\SI{}{\mwe}) or a high overburden of the order of \SI{100}{\mwe} or more, e.g. in specialized underground laboratories. 
The Dortmund Low Background Facility (DLB) combines the advantages of both approaches. 
The artificial overburden of \SI{10}{\mwe} already shields the hadronic component of cosmic rays. 
The inner shielding, featuring a state-of-the-art neutron shielding and an active muon veto, enables low-background gamma ray spectrometry at an easy-accessible location at the campus of the Technische Universität Dortmund.  

The integral background count rate between \SI{40}{\keV} and \SI{2700}{\keV} is \SI{2.528\pm0.004}{\counts\per\kilo\gram\per\minute}.
This enables activity measurements of primordial radionuclides in the range of some \SI{10}{\milli\becquerel\per\kilo\gram} within a week of measurement time.
\end{abstract}

\begin{keyword}
low-background gamma ray spectrometry \sep cosmic-muon veto \sep background-reduction techniques \sep HPGe detector
\end{keyword}

\end{frontmatter}


\section{Introduction}
\label{sec:introduction}

Modern gamma ray spectrometry is used in a variety of applications, ranging from monitoring of environmental samples to contamination control and material assaying for the usage in physics experiments. 
The latter is especially important for experiments built to search for rare phenomena like interactions of hypothetical WIMP particles in Dark Matter searches \cite{Cline:2014fua}, neutrinos in general or neutrinoless double beta-decay \cite{Barabash:2010ie}.
The signals in such experiments are expected in the range of \SI{1}{keV} to \SI{5}{MeV}, i.e. the energy range of nuclear reactions.
Furthermore, the expected event rates of the searched-for signals are as low as a few per day or even a few per year, depending on the physics case.
Consequently, the background contributions, in particular those due to the intrinsic contamination of the detector or shielding materials, have to be kept as low as possible. 
Residual contaminations have to be known precisely so that background contributions in physics runs can be determined, e.g. by using Monte-Carlo simulations.

One such ultra-low-background project is COBRA, a future experiment with the aim to search for neutrinoless double beta-decay in several different isotopes using CdZnTe semiconductor detectors \cite{COBRA.2015.2}. 
Since November 2013, the COBRA collaboration has operated a fully functional demonstrator with an array of 64 detectors at the Laboratori Nazionali del Gran Sasso (LNGS) underground laboratory \cite{COBRA.2015.1}. 
An important keystone of the R\&D phase of the COBRA experiment is the pre-selection of suitable materials based on in-depth material screening and radiopurity assays. 
Materials can be screened with very high sensitivity at ultra-low-background gamma ray spectrometry facilities such as the LNGS underground laboratory. 
By the rock overburden of 3800 meters of water equivalent (m w.e.) the muon flux is suppressed by six orders of magnitude \cite{borexino}. 
Such laboratories can reach sensitivities as low as \SI{10}{\micro\becquerel\per\kilo\gram} for the natural occurring radionuclides (uranium and thorium) with the disadvantage of rather limited availability, long measuring periods and high construction costs \cite{Heusser.2006.495}.

The Dortmund Low Background Facility (DLB) is a low-level gamma ray spectrometry laboratory designed for the detection of faint traces of radioactivity, and therefore enabling the screening of materials with a very high sensitivity with the initial motivation to support the COBRA collaboration. 
It is built above ground with an artificial overburden. 
Its high-purity germanium (HPGe) detector is installed in a multilayer inner shielding designed for surface operation and situated under nearly \SI{400}{\tonne} of overburden, corresponding to \SI{10}{\mwe} of coverage. 
Between the inner shielding and the artificial overburden, an active muon veto consisting of plastic scintillators is used to suppress the muon-induced contribution to the integral background by approximately one order of magnitude. 
Detection limits in the range of several \si{\milli\becquerel\per\kilo\gram} can be obtained for the primordial radionuclides from uranium and thorium.
The facility is thus suited for radiopurity assays and material pre-selection.

In Section 2, a brief overview of the sources of background radiation for gamma ray spectrometers is given and concepts how to shield against these are introduced.
The experimental setup, including the shielding concept and the readout chain of the gamma ray spectrometer, is described in Section 3. 
In Section 4 the spectroscopic properties of the detector system are presented.
Section 5 summarizes the background level obtained with the DLB and discusses the remaining peaks in the background spectrum. 
The detection limits, reached with the current setup, are also presented. 
A summary is given in Section 6. 

\section{Sources of background radiation and design considerations}
\label{sec:sources}

Background reduction is a crucial task for all gamma ray spectrometry laboratories aiming for sensitivities better than \SI{1}{\becquerel\per\kilo\gram}. 
The sources of background radiation for a low-level gamma ray spectrometer can roughly be categorized into two groups:
One are gamma rays from naturally occurring radioactivity including the radioimpurities of the detector setup and its environment, the other are airborne radon and cosmic ray induced gamma rays as well as direct interactions with atmospheric particles.

\subsection{Environmental radioactivity}

The major sources of environmental radioactivity are almost exclusively gamma rays from the \isotope[235]{U}, \isotope[238]{U} and \isotope[232]{Th} decay chains and \isotope[40]{K}. 
The starting nuclides of those decay chains as well as \isotope[40]{K} are primordial nuclides that have existed since the beginning of the stellar nucleosynthesis. 
They can be found in many minerals  and building materials. 
\autoref{tab:abundance} gives values for their element's natural abundance in the earth's crust \cite{CRC.2015} as well as the corresponding specific activities of the three most prominent sources for background (\isotope[238]{U}, \isotope[232]{Th} and \isotope[40]{K}). 
The concentrations vary for different materials, with the highest activities found in granites \cite{Heusser.1995.543}. 

\begin{table}
\caption[]{Abundances of the primorial radionuclides from uranium and thorium as well as potassium in the earth's crust. Abundances are taken from Ref. \cite{CRC.2015}. Specific activities are calculated with values from Ref. \cite{Firestone.1996.}.}
\label{tab:abundance}
\centering
\begin{tabular}{cSS}
\toprule %
\multicolumn{1}{c}{Element} 	& \multicolumn{1}{c}{Avg. abundance} 	& \multicolumn{1}{c}{Specific activity} 	\\
 						& [\si{\milli\gram\per\kilo\gram}] 		& [\si{\becquerel\per\kilo\gram}] 		\\
\midrule %
\text{Uranium}				&	2.7							&	33.6 \text{(\isotope[238]{U})}		\\
\text{Thorium}				&	9.6							&	39.0 \text{(\isotope[232]{Th})}		\\
\text{Potassium}			&	20900						&	663.9 \text{(\isotope[40]{K})}		\\
\bottomrule %
\end{tabular}
\end{table}

Although the surrounding of most spectrometers (e.g. concrete walls) can have rather high concentrations of primordial radionuclides, gamma rays from their decays can be shielded sufficiently by extra shielding of high-$Z$ materials. 
It is recommended to use \SIrange{10}{15}{\centi\metre} of lead to shield the detector's vicinity. 
This amount of lead is sufficient to decrease the intensity of \SI{3}{\MeV} gamma rays by about three orders of magnitude. 
Photons with lower energies are even more suppressed.
On the other hand, more lead can increase the background contribution from tertiary neutrons induced by cosmic ray muons via muon capture or by photonuclear reactions \cite{Gilmore.2008.}. 
\subsubsection{Airborne radon}
Secular equilibrium, i.e. the state in which the activities of all daughter nuclides are equal to the activity of their relative parents, is rarely achieved for the primordial decay chains.
Radon isotopes (especially \isotope[220]{Rn} and \isotope[222]{Rn}), for example, can escape from the matrix by recoil on ejection of the alpha particle or by diffusion, and thereby break the equilibrium of the decay chain within the matrix.

The radon isotope \isotope[222]{Rn}, originating from the \isotope[238]{U} decay chain, decays via \isotope[214]{Pb} (with \SI{351.9}{\keV} and \SI{295.2}{\keV} being the most prominent gamma rays) and \isotope[214]{Bi} (with \SI{609.3}{\keV}, \SI{1764.5}{\keV} and \SI{1120.3}{\keV} as most prominent gamma rays) with a half life of \SI{3.8}{\day} and is therefore more important in terms of background contribution for a gamma ray spectrometer. 
An average concentration in the air of about \SI{40}{\becquerel\per\cubic\metre} makes \isotope[222]{Rn} by far the strongest source of airborne radioactivity \cite{Heusser.1995.543}. 

The isotope \isotope[220]{Rn} from the \isotope[232]{Th} decay chain has a half life of \SI{55.6}{\second}. 
This sub-chain decays rather quickly via \isotope[212]{Pb} (\SI{238.6}{keV}) with a half life of \SI{10.6}{\hour} and \isotope[208]{Tl} (\SI{583.2}{keV} and \SI{2614.5}{keV}) with a half life of \SI{3.1}{\minute}.
Hence, the concentration of \isotope[220]{Rn} is below \SI{1}{\percent} of the \isotope[222]{Rn} concentration \cite{Heusser.1995.543}. 

Finally, \isotope[219]{Rn} from the \isotope[235]{U} decay chain is negligible for most low-level gamma ray spectrometers due to short half lives and less intense gamma rays in this sub-chain. 

Using filtrated air inside a gamma ray spectrometry laboratory can decrease the radon concentration.
To minimize the possibility of sample contamination with radon, the sample chamber is often flushed with gaseous nitrogen that is boiling out of the detector cooling dewar.
An air-tight sealing avoids the diffusion of radon through the detector shielding. 
Ultra-low-background facilities may even flush the sample preparation areas with gaseous nitrogen and store the sample under a protective atmosphere before the actual spectrometry measurement while radon is decaying. 
\subsubsection{Anthropogenic radionuclides} 
So-called anthropogenic radionuclides are introduced into nature by man.
Commonly found is \isotope[137]{Cs} which is brought into the environment due to nuclear accidents and nuclear weapon testing \cite{Gilmore.2008.}.
A commonly found contamination of steel is \isotope[60]{Co} which is induced during the industrial production process. 
The contamination is mainly coming from high-activity sources that were disposed as scrap and recycled into new steel products \cite{Koehler.2004}.

All construction materials need to be carefully assessed for contaminations before the are assembled in a low-level environment.
Lowest contaminations can be found in material that was produced many years ago and stored underground, avoiding the exposure to cosmic rays, causing activation in the material. 
Suitable examples are steel from the early twenties century or 2000-year-old roman lead, that can be found within sunken roman ships in the Mediterranean Sea or Black Sea. 

\subsection{Cosmic rays}

The remaining peaks in the background spectra of modern low-level germanium facilities are mainly induced by cosmic radiation. 

The primary cosmic rays reaching the earth from outside of our solar system (galactic cosmic rays) or directly from the sun (solar cosmic rays) are high energetic particles consisting mainly of protons and alpha particles. 
By interacting with nuclei of the outer atmosphere, they produce a variety of secondary cosmic ray particles that propagate towards the surface of the earth in hadronic and electromagnetic cascades.
 
The secondary cosmic rays are divided into two components, a soft component containing electrons, positrons and photons, and a hard component consisting of protons, neutrons and muons. 
The typical flux for cosmic muons with a mean energy of about \SI{3}{GeV} at sea-level is \SI{190}{\metre^{-2}\second^{-1}}. 
The relative fluxes of other particles compared to the muon flux are 0.34 for neutrons, 0.24 for electrons, 0.009 for protons and 0.0007 for pions \cite{Povinec.2008}.
 
Cosmic rays can show variations in energy and flux, both periodic and aperiodic. 
A possible influence is the atmospheric pressure if a denser atmosphere needs to be penetrated. 
That induces an annual modulation of the muon flux of about \SI[separate-uncertainty]{1.29(07)}{\percent} relative amplitude with an phase of \SI[separate-uncertainty]{179(6)}{\day} \cite{borexino}. \\
Extraterrestrial factors, like the magnetic modulation effect caused by solar flares, may have an impact of about \SI{10}{\percent}. 
Also the eleven year cycle of the sun activity can be observed in the flux variations of cosmic rays. 
This effect is caused by solar eruptions and inversely correlated with the sunspot activity cycle of the sun \cite{Niese2008209}. 
\subsubsection{Muon-induced background}
At sea level, cosmic rays mainly consist of muons. 
Due to the higher atmospheric depth at higher zenith angles, the particle flux decreases proportionally to $\cos^2 \left( \theta \right)$, where $\theta$ is the zenith angle. 
Cosmic muons can have energies of \SI{10}{GeV} or more, making them minimal ionizing particles with a stopping power of about \SI{2}{\MeV\per\gram\per\centi\metre\squared} \cite{Paul.1971}. 
Because of their high penetration power, the flux of cosmic muons can hardly be attenuated completely, e.g. \SI{1400}{\metre} of rock overburden (corresponding to about \SI{3800}{\mwe}) can reduce the particle flux by nearly six orders of magnitude to about \SI{0.0003}{\metre^{-2} \s^{-1}} \cite{borexino}.
However, relatively shallow overburdens can already reduce the muon flux by an useful amount \cite{Gilmore.2008.}. 
For shallow-depth laboratories, only muons and secondary neutrons from cosmic rays are relevant contributions in low-level measurements.

Muons can interact via several mechanisms and lead to various contributions to the background of a low-level gamma ray spectrum. 
Although direct interactions of muons with the germanium detector itself can lead to energy depositions of tens of MeV depending on the geometry of the detector, the main reason for background contributions of muons are electromagnetic processes \cite{Povinec.2008}. 
Because of the large amount of energy deposited in the germanium crystal by direct interactions, those events can be rejected rather easily from the spectrum by applying an energy discrimination.
Besides a continuous distribution of counts due to muon bremsstrahlung, often there is a dominant \SI{511}{keV} annihilation radiation peak in the spectrum originating from pair production or $\mu^+$ decays. 
The intensity of this peak depends on the material and the amount of shielding used. 

Another mechanism for cosmic muons to contribute to the background spectrum is via negative muon capture inside the high-$Z$ shielding material. 
As a consequence, gamma rays accompanied by fast neutrons are emitted by the de-exciting nucleus inside the shielding. 
\subsubsection{Neutron-induced background}
\label{sec:neutrons}
Neutrons have an attenuation length of about \SI{200}{\gram\per\centi\metre^2}. 
For an overburden of about \SI{10}{\mwe}, the contribution from neutrons originating from secondary cosmic rays are nearly negligible \cite{Heusser.1995.543}. 
Potentially remaining neutrons can be moderated and captured efficiently by a neutron absorber layer of the shielding. 
The protons of the nucleonic component are absorbed even by \SI{5}{\mwe} and are therefore negligible for the facility discussed in this paper as well.
Also, neutrons derived from uranium fission and $\left( \alpha,n \right)$ reactions are not considered here since they become important only below a few 100 meters of water equivalent of overburden. 
Hence, only tertiary neutrons are discussed in more detail here.

Tertiary neutrons dominate in setups with shieldings of a few meters of water equivalent. 
The capture of negative muons as well as photonuclear reactions and photofission of real and virtual photons associated with fast muons are strongly enhanced in high-$Z$ shielding material like lead. 
For negative muon capture, the atomic number of the nucleus is decreased by one and the excited nucleus emits gamma rays and neutrons when the structure is re-arranged. 
Ref. \cite{Heusser.1995.543} calculates the flux of neutrons produced by muon capture to be \SI{1.1e-4}{\centi\metre^{-2}\second^{-1}} for a muon flux of \SI{8e-3}{\centi\metre^{-2}\second^{-1}}. 
These fast neutrons can lead to activation and excitation of the shielding material and the detector itself by $\left(n, p\right), \left(n, \alpha\right), \left(n, 2n\right)$ and $\left(n, n' \gamma\right)$ reactions. 
\subsubsection{Cosmogenic activation}
During low-level operations, cosmogenic radionuclides are often clearly visible in the background spectrum at the beginning of an underground operation. 
Metastable nuclides (e.g. from capture reactions of isotopes of germanium or copper) or those in their excited states lead to several lines in the background spectrum. 
During production or storage above ground, cosmogenic radionuclides are produced in the shielding materials, e.g. copper or steel, the detector housing, e.g. aluminium, or the detector crystal itself. \\
Typical radionuclides that can be produced in copper used for the shielding or the detector housing are \isotope[54]{Mn}, \isotope[56]{Co}, \isotope[57]{Co}, \isotope[58]{Co}, \isotope[59]{Fe}, \isotope[60]{Co} or \isotope[65]{Zn} \cite{Brodzinski1990337}. 
The specific activities of theses isotopes can saturate at up to a few \si{\milli\becquerel\per\kilo\gram}, depending on the height above sea level during storage \cite{Laubenstein2009750}. 
In germanium, additionally to the aforementioned isotopes \isotope[55]{Fe}, \isotope[63]{Ni} or \isotope[68]{Ge} can be produced, all in the range of tens to hundreds of \si{\micro\becquerel\per\kilo\gram} at sea level \cite{Povinec.2008}.

After a rather long time of shielded operation, activation and decay of a specific isotope lead to a saturated level of activity, which strongly depends on the depth of the laboratory. 
This makes the cosmogenic radionuclides a crucial contribution to the background spectrum in ultra-low-background applications, often more important than intrinsic contaminations of the natural decay chains.
To avoid activation in the crucial parts of a spectrometer system, it is recommended to have the materials stored underground as soon as possible after their production. 
The production process, of course, needs to be optimized for background reduction as well. 
For further background reduction in deep underground laboratories,  it is even discussed to have germanium detector crystals grown underground \cite{Laubenstein.2004.167}.

Other cosmogenic radionuclides such as \isotope[3]{H}, \isotope[7]{Be}, \isotope[10]{Be}, \isotope[14]{C} or \isotope[36]{Cl} are negligible in most low-level gamma ray spectrometry applications and therefore not further discussed.

The interaction of fast neutrons with the different germanium isotopes leads to broadened peaks with asymmetric shapes to higher energies due to the nuclear recoil of the target nucleus, which is creating electron-hole pairs inside the detector.
Inelastic scattering via $\left(n,n'\gamma\right)$ reactions on \isotope[72]{Ge} and \isotope[74]{Ge}, for example, leads almost immediately (compared to the charge collection time) to a de-excitation of the nuclei with \SI{691.4}{\keV} and \SI{595.9}{\keV}, respectively \cite{Gilmore.2008.}. 
For similar events outside the germanium, only the de-excitation gamma is observed.

In contrast to inelastic neutron scattering in germanium, the interaction in materials around the detector crystal leads to sharp peaks in the spectrum since the recoil energy of the target nucleus remains in the material and only the emitted gamma ray has a certain chance to be detected in the germanium detector. Radionuclides that contribute to the background spectrum due to neutron interactions are \isotope[63]{Cu}, \isotope[65]{Cu}, \isotope[206]{Pb} or \isotope[207]{Pb} for example. 
 
Refs. \cite{Heusser.1995.543, Povinec.2008} recommend boron-loaded polyethylene (BPE) as a neutron moderator and absorber. 
For surface or shallow-depth laboratories, it is recommended to place a rather thick neutron shield inside the lead shielding to stop tertiary neutrons. 
However, placing the moderator inside the lead leads to a larger mass for muon interaction and a cubic increase in cost for the shielding.

Placing an active veto shield outside the lead shielding is an effective technique for background suppression of neutron events, as long as the veto shield comes with a high detection probability for the penetrating cosmic muons.
Organic scintillators are sensitive to muon interactions as well as to neutron penetration and can be used for anti-coincidence measurements with the germanium detector. 

At deep underground laboratories, e.g. the LNGS, where the overall neutron flux is not dominated by cosmic rays but by neutrons from uranium fission and $\left( \alpha,n \right)$ reactions, the neutron shield is placed best outside to shield against those neutrons from uranium and thorium, that is contained in the walls of the laboratory. 

\section{Experimental setup}
\label{sec:setup}

The DLB is set up above ground and installed in the experimental hall of the Department of Physics of the Technische Universität Dortmund. 
Unlike most ultra-low-background laboratories, an on-surface location was chosen to provide easy access as well as the possibility to use the facility in education of physics students. 

\begin{figure}[h!]
\includegraphics[width=0.99\columnwidth, angle=0]{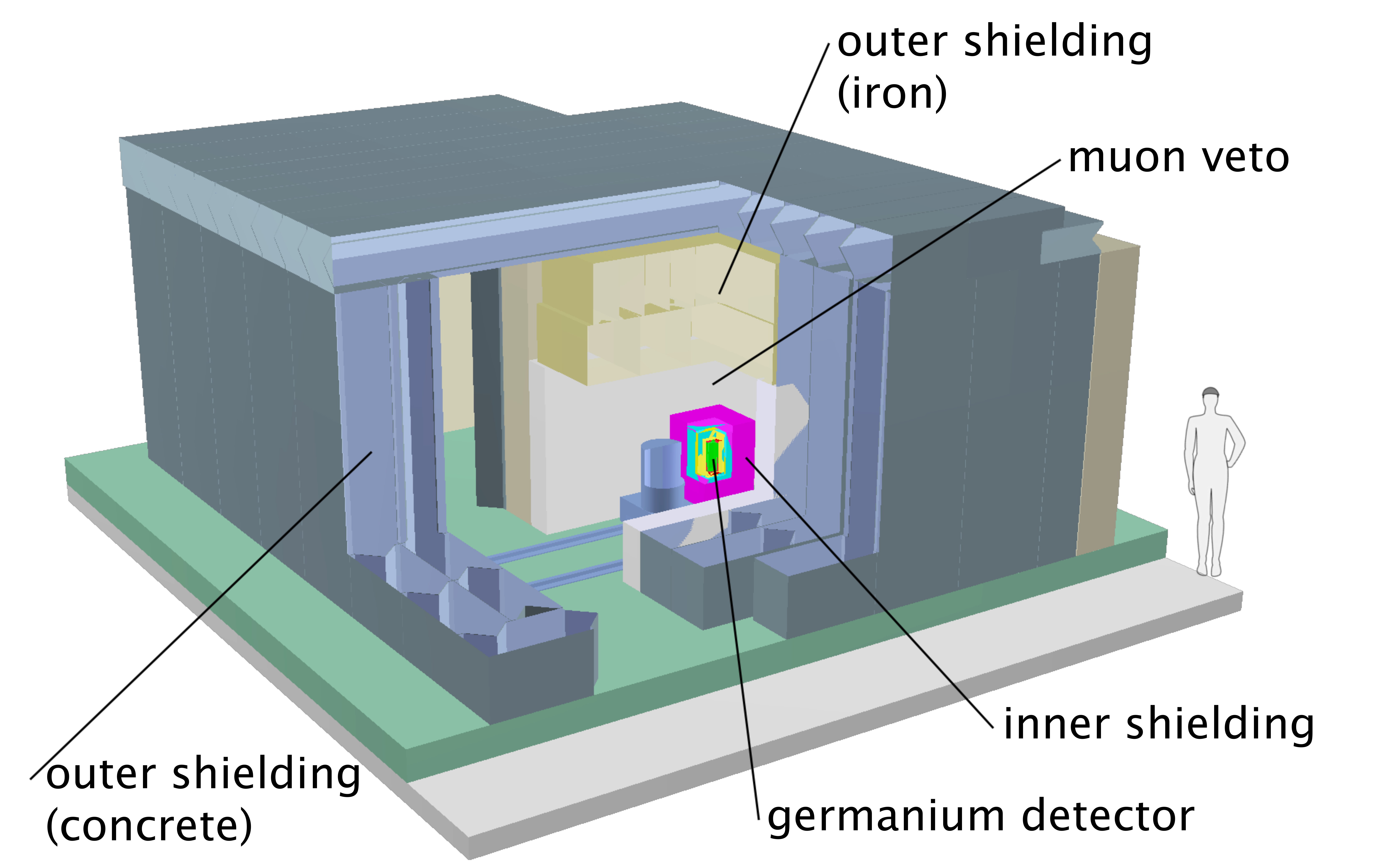}
\caption[]{The setup of the DLB as a (simplified) digital model. The outer shielding consists of iron and concrete blocks. The inner shielding of lead, boron-loaded polycarbonate and copper houses the HPGe detector. In between both, the muon veto is installed.}
\label{fig:outer_shield}
\end{figure}

The four major components of the laboratory are the high-purity ultra-low-background (ULB) germanium detector, the inner shielding, an active muon veto and the outer shielding. 
The inner shielding, also referred to as the lead castle, has a multilayer design and shields against environmental radioactivity and neutrons.  
Plastic scintillators read out by photomultiplier tubes (PMT) are installed, partly covering the inner shielding, to detect cosmic particles (muons as well as potentially secondary neutrons) that penetrate the outer shielding and hit the inner shielding. 
Events that are caused directly or delayed by such particles are not recorded in the measured spectra of the germanium detector. 
The outer shielding serves as an overburden that causes an attenuation of cosmic rays and houses the entire laboratory, including readout electronics and sample preparation.
\autoref{fig:outer_shield} shows the DLB setup including the four major components, that will be discussed in more detail in the following sections.

\subsection{Germanium detector and sample chamber}

The main component of the DLB is its HP germanium detector in a ULB version manufactured by Canberra Semiconductor NV in Olen, Belgium. 
The standard electrode ($p$-type) semi-coaxial diode has a mass of \SI[separate-uncertainty]{1.247(2)}{\kg} with dimensions of \SI{71}{mm} in diameter and \SI{66}{mm} in length resulting in a volume of \SI{234}{cm^3} corresponding to a relative efficiency of \SI{60}{\percent} and a peak-to-Compton ratio of 77:1. 
At \SI{1332.5}{keV} (\isotope[60]{Co}), the energy resolution quoted by the manufacturer is \SI{1.84}{keV} (\SI{0.14}{\percent}) \cite{CanberraIndustries.2009.SEGe, CanberraIndustries.2010.drawing}. 
\subsubsection{Detector front window sensitivity scans}
In addition to the parameters describing the germanium crystal quoted by the manufacturer, several measurements were performed to characterize the detector and its geometry accurately. 
Detailed information about the detector crystal are necessary for reliable Monte Carlo simulations, that are used for the full-energy detection efficiency determination. 

The scanning of the front window of the sensitive crystal volume was performed after the installation of the detector inside of the setup described in this paper. 
The dimensions of the sensitive crystal front window of the detector are determined by using a calibrated \SI{40}{MBq} \isotope[241]{Am} source inside a custom-made collimator consisting of copper, lead and poly-methyl-methacrylate (PMMA).
It can be placed on a positioning plate with precisely known positioning points across the detector end-cap.
By using the \SI{59.54}{keV} gamma rays as well as the peaks at \SI{98.97}{keV} and \SI{102.98}{keV} of \isotope[241]{Am}, an extensive study was performed to determine the exact position of the crystal as well as the homogeneity of the detector's front dead layer. 
The term dead layer refers to the outer electrode contact, which is not contributing to the sensitive crystal volume. 
The local thickness of the dead layer can be determined for each measuring point by calculating the ratio between the two peak areas. 
The number of counts in the peak areas depends on the absorption coefficients between the source and the sensitive volume of the detector. 
The average count rate at \SI{59.54}{keV} across the detector surface deviates by \SI{-20}{\percent} to \SI[retain-explicit-plus]{+11}{\percent} of the overall mean count rate.
Assuming that only the thickness of the dead-layer is causing this effect, the variation of the dead layer thickness is estimated to be about \SI{300}{\micro\metre}.
This value is comparable to variations found in other detectors \cite{neddermann_2014_phd}. 

The overall dead-layer thickness was determined by using a calibrated \isotope[133]{Ba} source. 
The determination bases on the energy dependent absorption coefficient of the material between the source and the sensitive volume of the detector. 
In the measured spectrum, the ratios of the peak areas (here, gamma lines at \SI{53.2}{\keV}, \SI{79.6}{\keV}, \SI{81.0}{\keV} and \SI{160.6}{\keV}) are determined and represent the target values that have to be reproduced in Monte-Carlo simulation.
The weighted average for the effective dead-layer thickness, combining the three different ratios, is calculated to be \SI{437(16)}{\micro\metre}, which is a more precise value than the aforementioned estimation \cite{neddermann_2014_phd}.
This value is only \SI{62}{\percent} of the nominal thickness, given by the manufacturer, which is probably due to the continuous cooling of the detector crystal, that is preventing \isotope[]{Li} atoms from diffusing into the crystal \cite{CanberraIndustries.2010.drawing}.
\subsubsection{Sample chamber and detector holder}
Permanent cooling with liquid nitrogen is provided by a ULB U-style cryostat which prevents any direct paths for radiation into the shielding. 
Except for two controlled thermal cycles, the detector has been kept permanently cooled since its delivery in 2008 to suppress slow diffusion of the lithium atoms of the outer contact into the crystal.
The detector is mounted inside a holder made of \SI{99.99}{\percent} pure copper within an end-cap made of radiopure aluminium with a uranium and thorium content of less than 1 part per billion (\SI{}{ppb}) \cite{CanberraIndustries.2008.ULB}. 
The aluminium end-cap and the thickness of the outer contact lead to a typical energy threshold of about \SI{40}{keV} and low sensitivity to energies marginally above. 
Directly above the germanium detector is the sample chamber of approximately \SI[product-units =  brackets-power]{94 x 94 x 93}{\milli\m} size. 

\subsection{Outer and inner shielding}

The artificial overburden consists of \SI{1}{\m} to \SI{1.5}{\m} barite concrete, adding up to a mass of $\SI{319}{t}$, and a \SI{1.17}{\metre} thick cast iron ceiling with a weight of $\SI{43}{t}$ forming an \SI{3}{\metre} $\times$ \SI{1.4}{\metre} $\times$ \SI{1.7}{\metre} ($l \times w \times h$) measuring tunnel and a control room (see \autoref{fig:outer_shield}). 

\begin{figure}[h!]
\includegraphics[width=0.99\columnwidth, angle=0]{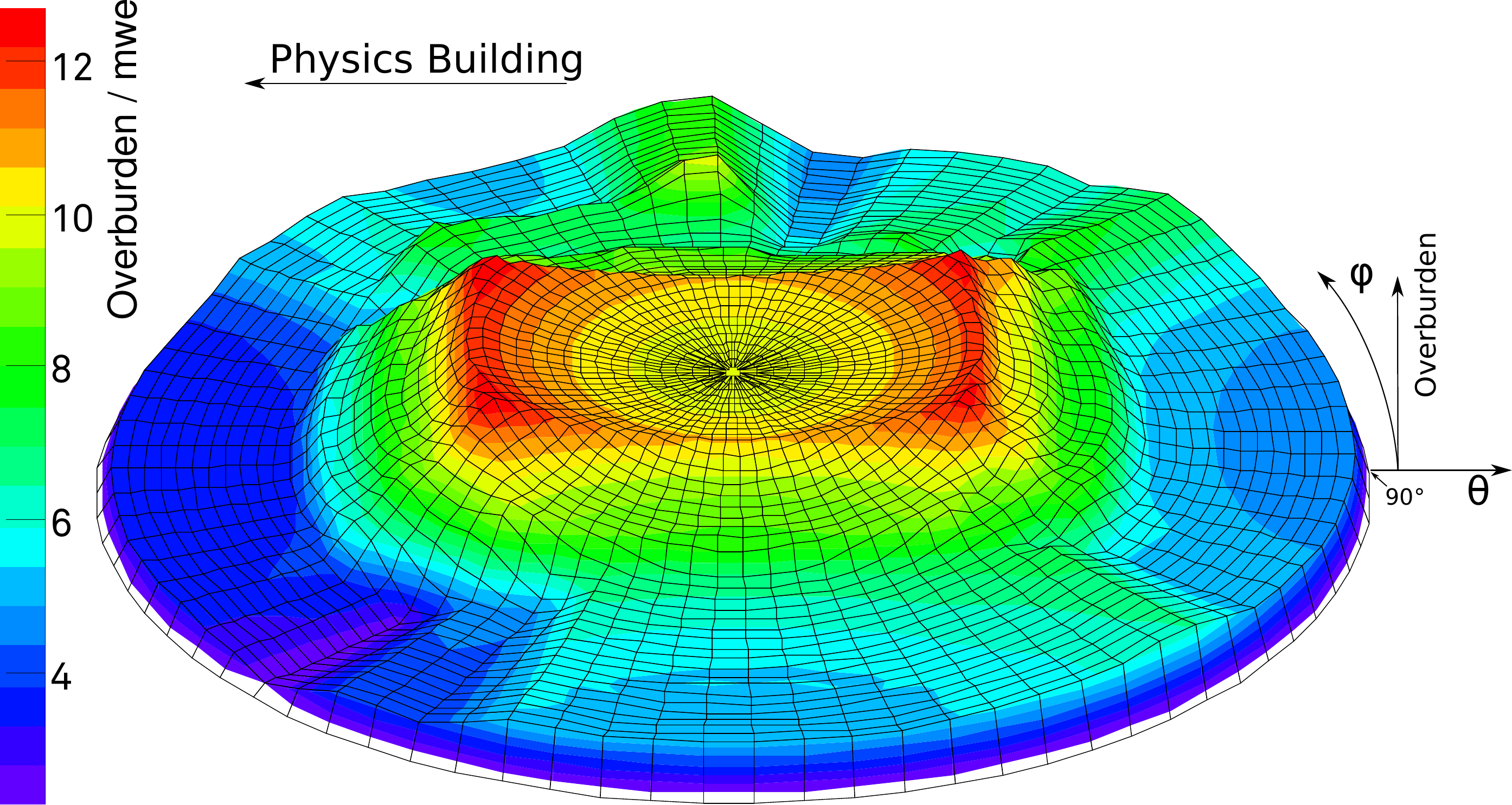}
\caption[]{Angular distribtion of the overburden of the DLB. The zenith angle $\theta$ corresponds to the radius, the height $z$ and the additional color coding to the overburden in meters of water equivalent. The influence of the cuboid-shaped iron block is clearly visible \cite{neddermann_2014_phd}.}
\label{fig:overburden}
\end{figure}

The overburden corresponding to the outer shielding of the DLB can be illustrated by an angular distribution.
For each direction in the upper hemisphere, the overburden is calculated analytically in meters of water equivalent by integrating over the density of the material in \SI{10}{\centi\metre} steps along the way.
\autoref{fig:overburden} shows the overburden of the upper hemisphere around the germanium detector up to a distance of \SI{10}{\metre}.
It is at least \SI{10}{\mwe} for a zenith angle $\theta$ of less than \SI{45}{\degree} measured from the germanium-detector position. 

Since the muon flux decreases approximately as $\cos^2\theta$ for muons with an energy of approximately \SI{3}{\GeV}, the overburden becomes less important for larger angles. 
In its present form, the overburden shields the soft component of cosmic rays completely and strongly reduces the hadronic component. 
Neglecting the remaining neutron flux, which is suppressed by about four orders of magnitude, cosmic muons are the only particles able to penetrate the overburden.
Ref. \cite{Gilmore.2008.} states that \SI{10}{\mwe} are enough to suppress the muon flux by \SI{60}{\percent}.
An approximated attenuation of cosmic muons is given by

\begin{equation}
a_{\mu} \left( m_{o} \right) = 10^{-1.32 \log d - 0.26 \left( \log d \right)^2}
\end{equation}

where $d = 1+ m_{o}/10$ and $m_{o}$ is the given overburden in meters of water equivalent \cite{Theodorsson.1996.}. 
With an overburden of \SI{10}{\mwe} the relative attenuation of the cosmic muon flux is
\begin{equation}
I_{\mu} = \left( 1- a_{\mu} \left( 10\right)\right) = \left(1-0.38\right) = 0.62,
\end{equation}
which is in good agreement with the aforementioned value and actual measurements performed inside the DLB shielding. 
If measured with a two-layer scintillator setup in coincidence, the muon flux at the position of the germanium detector can be determined to be \SI[separate-uncertainty]{75.45(05)}{\per\metre\squared\per\second}.
Compared to the unsuppressed flux of about \SI{190}{\per\metre\squared\per\second}, this gives a relative attenuation of about \SI{60.5}{\percent}. 
The minimal energy of muons passing vertically through the overburden and reaching the muon veto is calculated to be about \SI{1.56}{\GeV}.

\begin{figure}[h!]
\includegraphics[width=0.99\columnwidth, angle=0]{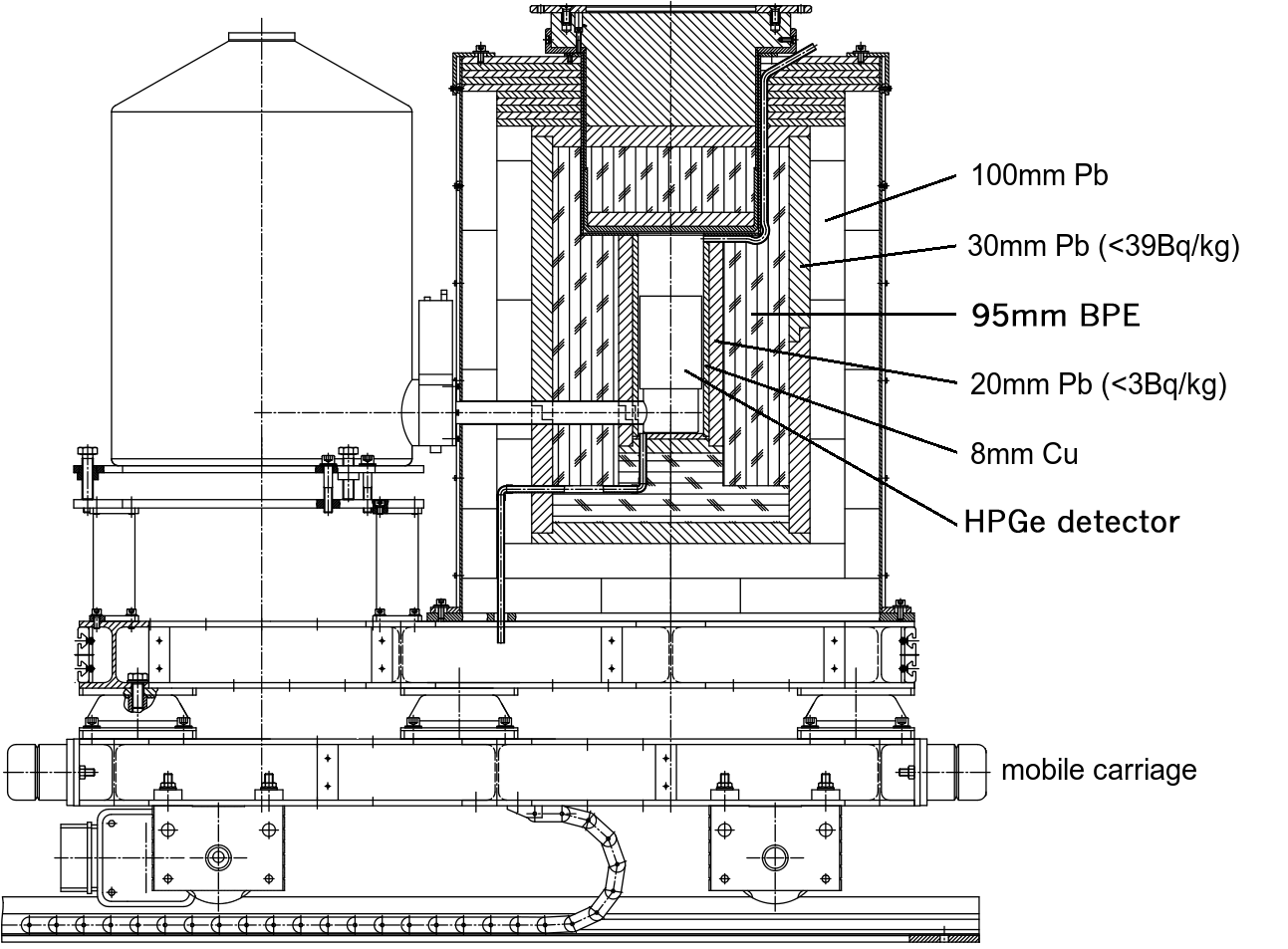}
\caption[]{Schematic drawing of the different layers of the shielding. Beginning by the outer most layer: \SI{100}{mm} standard lead, \SI{30}{mm} lead with an activity of less then \SI{39}{\becquerel\per\kg} \isotope[210]{Pb}, \SI{95}{mm} boron loaded polyethylene, \SI{20}{mm}  lead with an activity of less then \SI{39}{\becquerel\per\kg} \isotope[210]{Pb} and \SI{8}{mm} electrolytically produced copper. The whole setup, including the massive shielding and the nitrogen dewar, is mounted on a wagon that can be moved with an electrically powered engine.}
\label{fig:inner_setup}
\end{figure}

To shield the germanium detector against environmental radioactivity, and secondary as well as tertiary neutrons, the DLB features a multilayer inner shielding, especially designed for above-ground operation. 
The inner shielding is based on the shielding concept developed in Ref. \cite{Isus}. 
Its outer dimensions are \SI{0.6}{\metre} $\times$ \SI{0.6}{\metre} $\times$ \SI{0.8}{\metre}.
The entire inner setup has a mass of \SI{2.8}{\tonne}.
As shown in \autoref{fig:inner_setup} the shielding consists of five different layers. 
These are from the outermost to the innermost:
\begin{itemize}
	\item \SI{100}{\milli\metre} standard lead in form of bricks of size \SI{200 x 100 x 50}{\milli\metre}, placed in two layers to avoid gaps. They have an unknown activity of \isotope[210]{Pb},
	\item \SI{30}{\milli\metre} lead with an activity of less than \SI{39}{\becquerel\per\kg} \isotope[210]{Pb},
	\item \SI{95}{\milli\metre} boron-loaded polyethylene (BPE) as a neutron moderator and absorber,
	\item \SI{20}{\milli\metre} lead with an activity of less than \SI{3}{\becquerel\per\kg} \isotope[210]{Pb} and
	\item \SI{8}{\milli\metre} electrolytically produced copper.
\end{itemize}
As discussed before, the shielding consists of no more than \SI{15}{\centi\metre} of lead to avoid an enhancement of tertiary neutron production.  

For an easy access to the sample chamber without corrupting the shielding, a removable plug is installed on the top of the lead castle consisting of the same multilayer layout. 
Only the thicknesses of the outermost lead layer and the copper pot enclosing the layers differ slightly from the structural design.
The plug can be moved with a crane which is installed in the laboratory. 

The neutron shield is placed inside the massive shielding to minimize the effect of neutron interactions with copper and accompanying activation processes.
Hence, most of the high-Z material is located outside, and tertiary neutrons induced by muons penetrating through the lead are moderated and finally stopped in large part within the BPE layer. 
The material used contains about \SI{14.4}{wt\percent} hydrogen \cite{Wieser.2009.2131} and is therefore a good neutron moderator. 
By mixing diboron trioxide into the material, a boron content of about \SI{2.7}{wt\percent} of the BPE is reached. 
Since the neutron shield needs to be positioned close to the detector, it needs to be adequately radiopure. 
The screening measurements, shown in \autoref{tab:bpe_activity}, confirm the radiopurity of the used material \cite{neddermann_2014_phd}.

\begin{table}[tb]
\caption[Radiopurity data of BPE.]{Radiopurity data of boron-loaded polyethylene (sample mass \SI{249.9}{\gram}, live-time \SI{340512}{\second}), measured at LNGS (Detector GePaolo) by M. Laubenstein \cite{neddermann_2014_phd}.}
\label{tab:bpe_activity}
\centering
\begin{tabular}{ccS[separate-uncertainty, table-comparator = true]}
\toprule
Chain & Nuclide  									& {Activity [\si{\milli\becquerel\per\kg}]	}	\\
\midrule
\multirow{2}{*}{\isotope[232]{Th}}    	& \isotope[228]{Ra}  		& <3.0\\ 
                                				& \isotope[228]{Th}  		& 7.0+-2.7\\
\midrule
\multirow{3}{*}{\isotope[238]{U}}    	& \isotope[226]{Ra}  		& 12+-3\\ 
                                				& \isotope[234]{Th}  		& <75\\            
	                    				& \isotope[234m]{Pa} 	& <230\\
\midrule
                                				& \isotope[235]{U}     	& <9.8\\
                                				& \isotope[40]{K}    		& <64\\
                                				& \isotope[137]{Cs}  		& <4\\
\bottomrule
\end{tabular}
\end{table}

Between the BPE and the germanium detector, an ultra-low-activity layer of lead is placed to absorb the gamma rays that can be emitted during the neutron capture of boron. 
In \SI{93.7}{\percent} of the $\isotope[10]{B}\left(n,\alpha\right)\isotope[7]{Li}$ reactions, a gamma ray of \SI{477.6}{keV} is emitted by the excited \isotope[7]{Li}. 
The attenuation of the gamma ray intensity due to the innermost lead layer is analytically calculated to be \SI{98.1}{\percent}.

Details on the cleaning procedures for all parts and of the assembly can be found in Ref. \cite{neddermann_2014_phd} and references therein. 

\autoref{fig:inner_setup} shows the complete inner setup, including the massive shielding and the nitrogen dewar.
The setup is mounted on a vibration-damped heavy-load wagon and can be moved on rails with an electrically powered engine. 
To achieve maximum angular coverage of the overburden, the wagon can be moved under the cast iron that forms a tunnel. 
Consequently, the setup is moved out of the tunnel to introduce or change samples and refill the nitrogen reservoir. 

The inner shielding is surrounded by steel sheets which provide a sealing and an encapsulation of the system against radioactivity from radon in the air. 
Airborne radon is flushed out of the measuring chamber by constant evaporation of gaseous nitrogen from the detector dewar. 
The volume is exchanged completely in less than \SI{3}{\minute}.
Investigation of the radon concentration in the air inside the laboratory showed a relatively low activity of \SI[separate-uncertainty]{32.1(10)}{\becquerel\per\cubic\metre} compared to the aforementioned mean value of \SI{40}{\becquerel\per\cubic\metre} of \isotope[222]{Rn}. 
Although the concentration is even lower when using fresh air from outside, it is chosen to use air from the inside of the experimental hall.
Hence, the volume of the hall is used as a thermal buffer against the environment. 
Consequently, the radon concentration inside the laboratory is lowered to \SI[separate-uncertainty]{20.9(5)}{\becquerel\per\cubic\metre}, using air from the experimental hall. 
A slight overpressure inside the laboratory avoids dust diffusion into the working space. 
In addition, the influent air is filtered for minimization of dust inflow.  
However, peaks due to \isotope[222]{Rn} and its progenies can still be found inside the background spectrum since the sample chamber is opened regularly for sample changes.

\subsection{Active muon veto}
\label{sec:veto}

For shallow-depth laboratories the main background is induced by cosmic rays that penetrate the overburden and interact with the shielding material or the detector itself. 
This background component is lowered by using an active muon veto.

The muon veto consists of organic plastic scintillators, which have a high photon yield and a fast signal rise time of a few nanoseconds.
To obtain the best possible coverage, the detectors are placed close to the lead castle underneath the artificial overburden. 
Since it is placed outside the inner lead shielding, radiopurity is not crucial for the scintillation material, the photomultiplier tubes or the mounting parts. 
In different upgrade stages, the veto has been built in walls covering all sides of the setup. 
At the current stage, there are four veto walls installed, covering the top and three sides of the inner shielding. 
The last uncovered side is facing the nitrogen dewar, due to the more complicated geometrical structure. 
However, both the top plane and the side walls are extended towards the dewar to minimize the uncovered solid angle in this direction. 
Although the contribution originating from this side is considered to be rather small it will be closed during the next upgrade stage. 

Due to the stopping power of \SI{2}{\MeV\per\gram\per\centi\metre\squared} for muons in plastic scintillators, a thickness of at least \SI{5}{cm} for each scintillator is recommended to enable sufficient discrimination between muons and environmental gamma radiation by applying pulse height discrimination. 
For the muon veto described here, scintillator plates of \SI{5}{\milli\metre} thickness are used instead. 
The active area of each single scintillator in the top and side planes is \SI{122 x 20}{\centi\metre} and for the backside scintillators \SI{107 x 20}{\centi\metre}, respectively. 
 
\begin{figure}[h!]
\includegraphics[width=0.99\columnwidth, angle=0]{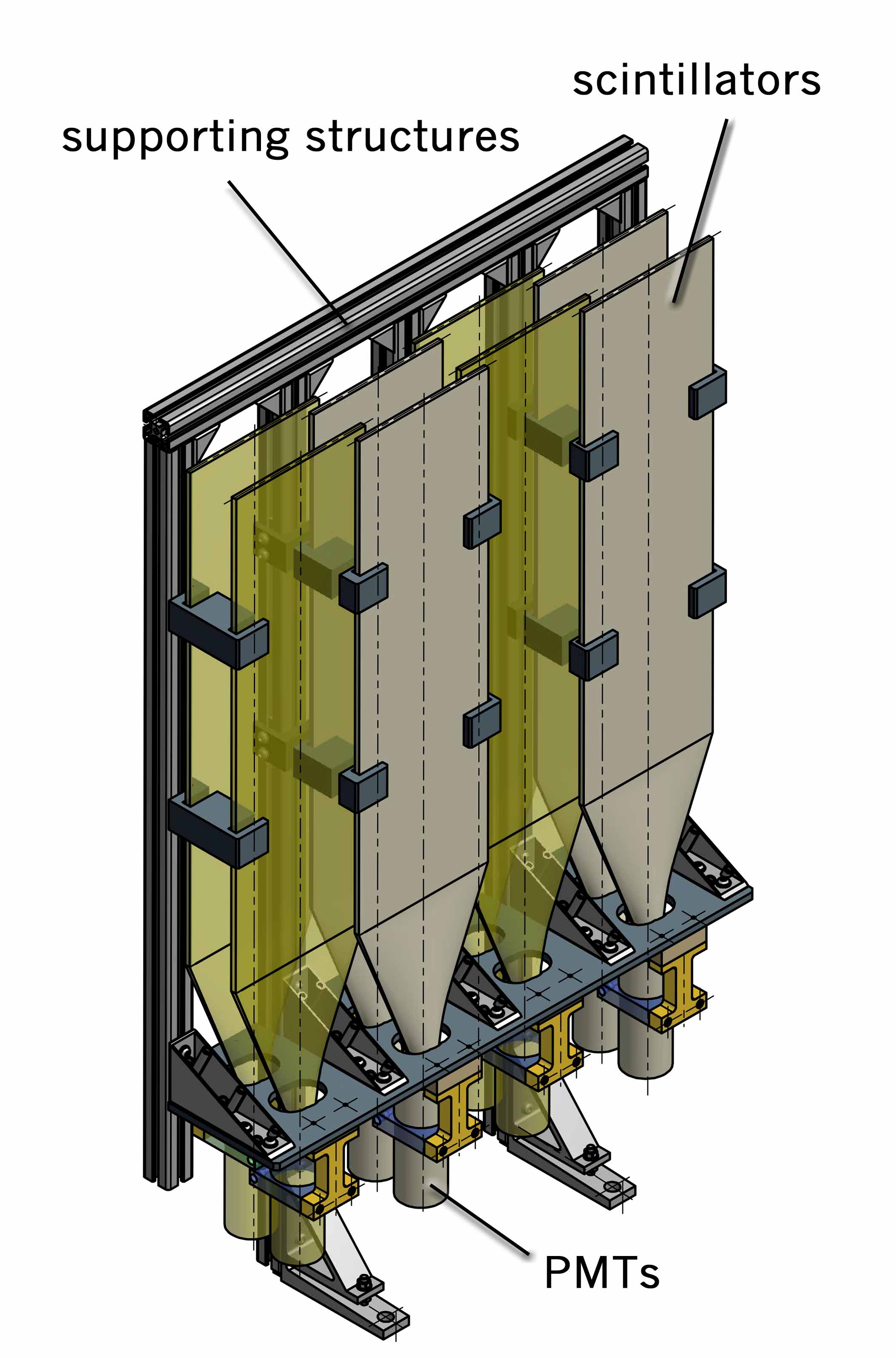}
\caption[]{Digital model of the backside plane of the muon veto, indicating the interlacing setup of the two scintillaton detectors each for coincidence measurements. The supporting structure is made of aluminium.}
\label{fig:backside}
\end{figure}

The muon veto makes use of a setup with two scintillators covering the same area of the solid angle, seen by the germanium detector. 
The interlacing setup is shown in \autoref{fig:backside} for the backside plane of the muon veto.
If a signal above the threshold is given by two scintillators at the same time, it is interpreted as cosmic ray particle, most likely a muon. 
Gamma rays interacting in both scintillators are very unlikely and therefore negligible. 
Additionally, randomly generated coincidences, produced by noise within the photo multiplier tubes, are negligible.
For the typical count rates produced by two scintillators with the used active areas, the signal contribution due to noise is well below \SI{0.1}{\percent}. 

This two-scintillator setup results in a muon detection efficiency comparable to a single thick scintillator.
Testing measurements result in a muon detection efficiency of \SI{99.5\pm0.2}{\percent} for a two-scintillator setup. 
This setup gives values that are in very good agreement with the literature values for the cosmic muon flux.

The used scintillators are Bicron \textit{BC-408} \cite{Crystals2009}, which are read out by photomultiplier tubes.
The wavelength of maximal emission of the scintillation light is equal to the wavelength of maximal sensitivity of the Hamamatsu \textit{R2490-05} photomultiplier tubes, which is \SI[separate-uncertainty]{420(50)}{\nano\metre} \cite{Photonics1992}.

For each scintillator, a photomultiplier was found to achieve sufficient signal output.
Within characterization measurements, a threshold was determined for each detector for optimal noise suppression.  
The NIM pulses generated by the discriminators are logically combined by coincidence modules representing the interlacing setup. 
After adding up all contributions from different parts of the detector, an overall signal, representing the complete muon veto, is generated and applied to the gate input of the multi-channel analyzer of the germanium detector. 

After installation in the DLB setup, the veto count rate of the complete muon veto was measured to be \SI[separate-uncertainty]{275(1)}{\counts\per\second}. 
Compared to the calculated suppressed muon flux of about \SI{200}{\counts\per\second} through the covered area, this gives an effectiveness of \SI{73}{\percent} for the muon veto. 
The counts that are not induced by cosmic muons are caused by the insufficient discrimination between cosmic muons and environmental gamma radiation in the scintillators covering the edges of the setup, where no coincidence techniques can be applied. 
However, the overall dead time of the currently operated system is well below \SI{1.5}{\percent}. 
By applying the active muon veto to the germanium gamma ray spectrometry measurements, the integral background count rate is lowered by nearly an order of magnitude. 

\subsection{Data acquisition electronics}

\begin{figure}[h!]
\includegraphics[width=0.99\columnwidth, angle=0]{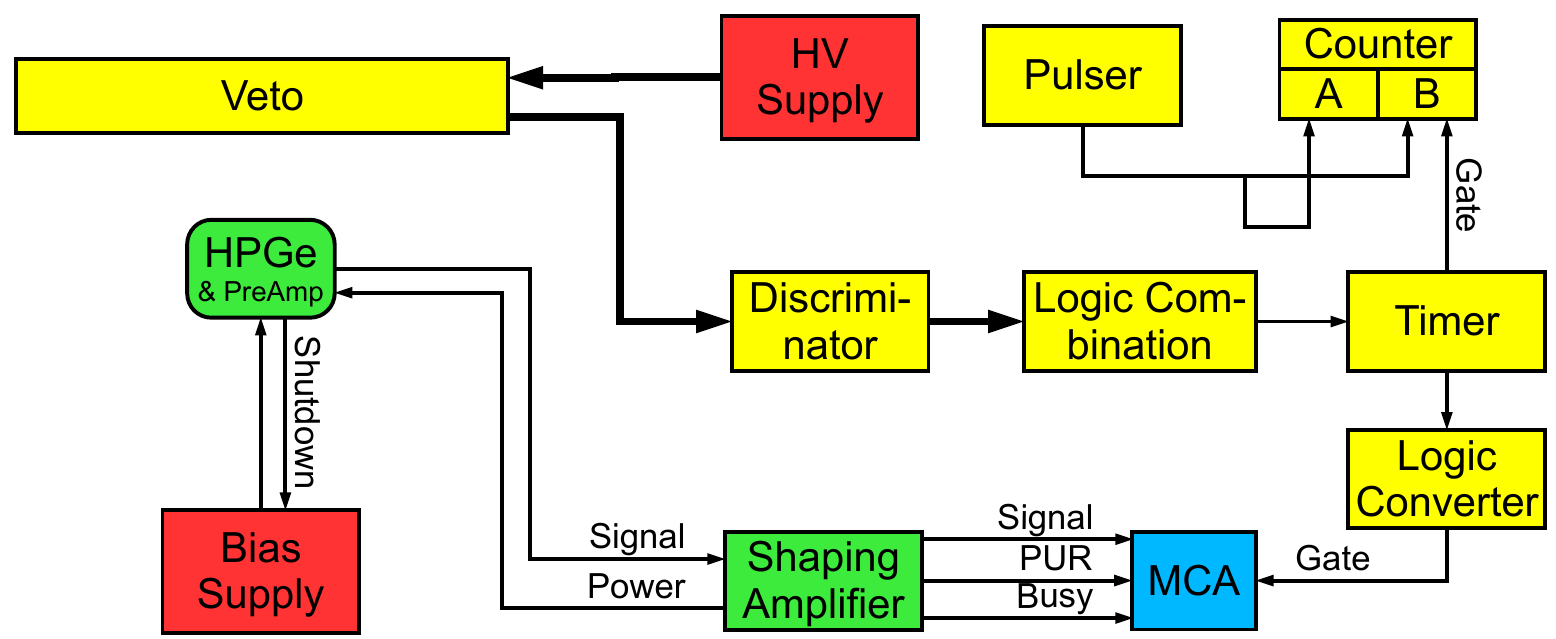}
\caption[]{Data acquisition electronics with conventional read out. Bold arrows indicate multiple connections. The logic combination module represents several units (concidence, multiplicity logic and OR units), that are combining the discriminated signals coming from the PMTs so that the two-layer coverage is assured \cite{neddermann_2014_phd}.}
\label{fig:readout}
\end{figure}

\autoref{fig:readout} gives a schematic of the data acquisition electronics (DAQ) used in the DLB.
The raw PMT signals are converted into fast negative NIM logic outputs by several discriminator units. 
Each threshold, adjustable per channel or unit, is chosen to match the amplification characteristics of the individual PMT. 
The signals are then checked for coincidences within overlaying scintillators. 
The main parts of the top and side planes of the muon veto are checked for coincident signals with a multiplicity logic unit. 
Since one layer of scintillators is readout by two PMTs each, the logic requirement for this part is that three input signals have to be coincident. 
However, the outer edges and the backside plane are single-readout scintillators and hence, one coincidence module is sufficient for overlapping detectors. 
Lastly, the coincident signals are combined with several OR units to one single veto signal. 

For each veto signal given by the complete muon veto, an adjustable time window is applied to the multi-channel analyzer's gate input, rejecting signals in the germanium detector for this duration. 
Typical values of \SI{2}{\micro\second} up to almost a millisecond can be found at other detector systems \cite{neddermann_2014_phd}. 
A study was performed to find the optimal time window, that is rejecting signals due to prompt and delayed processes induced by muons without losing too much lifetime of the measurement. 
Consequently, a time window of \SI{50}{\micro\second} is chosen, assuring high efficiency but with a highly reduced dead time of about \SI{1.35}{\percent}, corresponding to the aforementioned overall veto count rate.

In order to determine the relative dead time due to the muon veto, a counter is gated by the veto signal while counting pulses from a pulser.
This number is then divided by the total number of pulses applied.

\section{Spectroscopic properties of the DLB}
\label{sec:spec_prop}

In the following, the spectroscopic properties of the DLB's germanium detector are discussed, such as the energy calibration and the energy resolution. 
Furthermore, the determination of the full-energy peak (FEP) detection efficiency using Monte Carlo (MC) simulations for individual sample geometries is briefly described.

\subsection{Energy calibration and resolution}

The germanium detector is regularly calibrated using a \isotope[232]{Th} source placed inside the sample chamber. 
After acquiring a sufficiently large number of events in all relevant peaks of the spectrum, an automated function is used for peak search and identification: each peak is fitted with a Gaussian plus linear function, and the side bands are used to estimate the background level for this energy interval.

\begin{figure}[h!]
\includegraphics[width=1.1\columnwidth, angle=0]{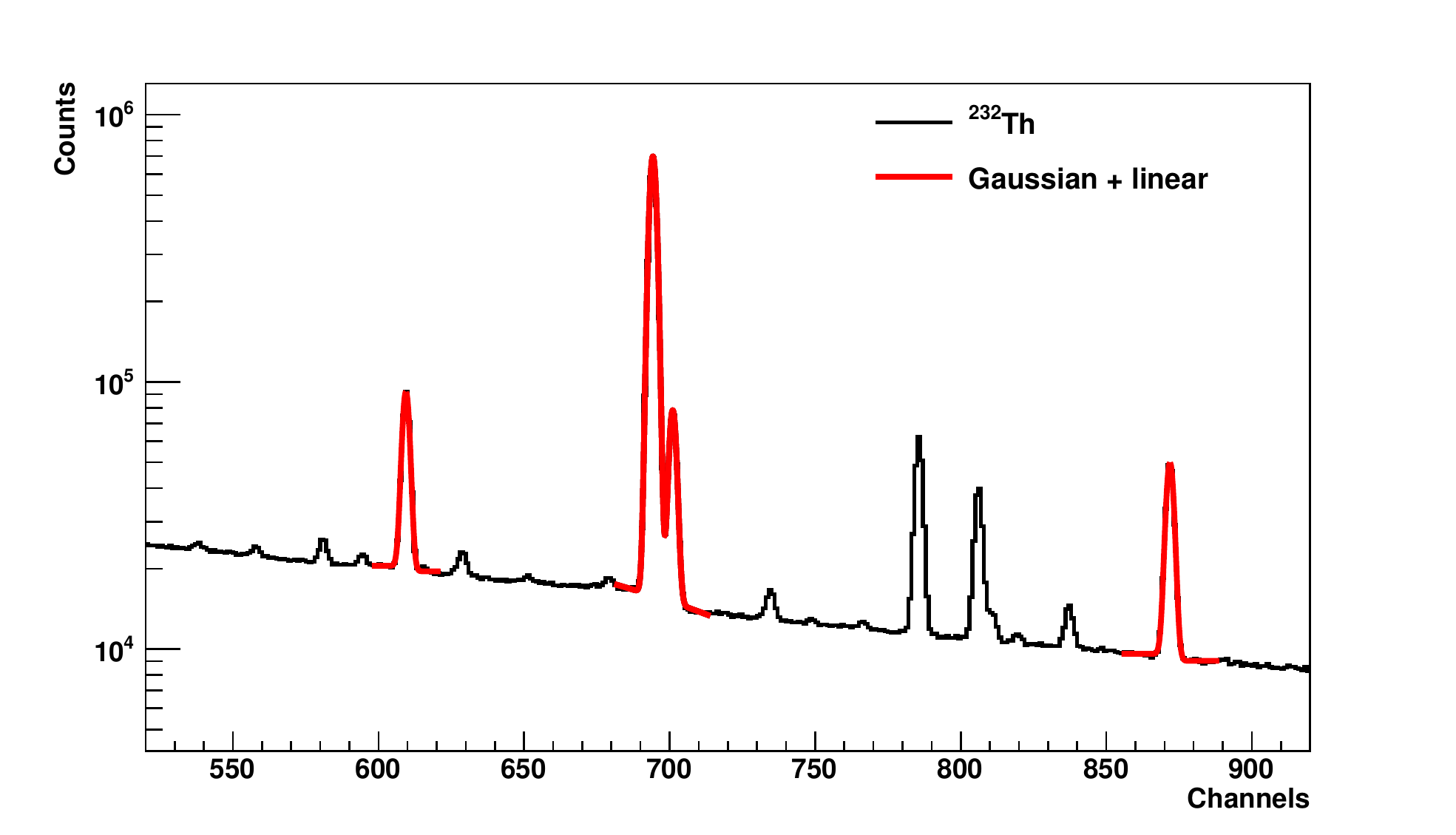}
\caption[]{Partial energy spectrum (black line) of the \isotope[232]{Th} calibration source with fitted Gaussian plus linear functions (red lines). The four gamma lines from fitted in this sector are \SI{209.25}{\keV} (\isotope[228]{Th}), \SI{238.63}{\keV} (\isotope[212]{Bi}), \SI{240.99}{\keV} (\isotope[220]{Rn}) and \SI{300.09}{\keV} (\isotope[212]{Bi}).}
\label{fig:th232}
\end{figure}

\autoref{fig:th232} shows several gamma peaks, measured using the \isotope[232]{Th} calibration source.
The gamma lines in this energy interval which are considered for the calibration are \SI{209.25}{\keV} (\isotope[228]{Th}), \SI{238.63}{\keV} (\isotope[212]{Bi}), \SI{240.99}{\keV} (\isotope[220]{Rn}) and \SI{300.09}{\keV} (\isotope[212]{Bi}).
The mean of each Gaussian function is used for the linear calibration function.
The maximum deviation between the measured energy deposition within the detector and the known peak energies is well below \SI{0.02}{\percent}.

\begin{figure}[h!]
\includegraphics[width=1.1\columnwidth, angle=0]{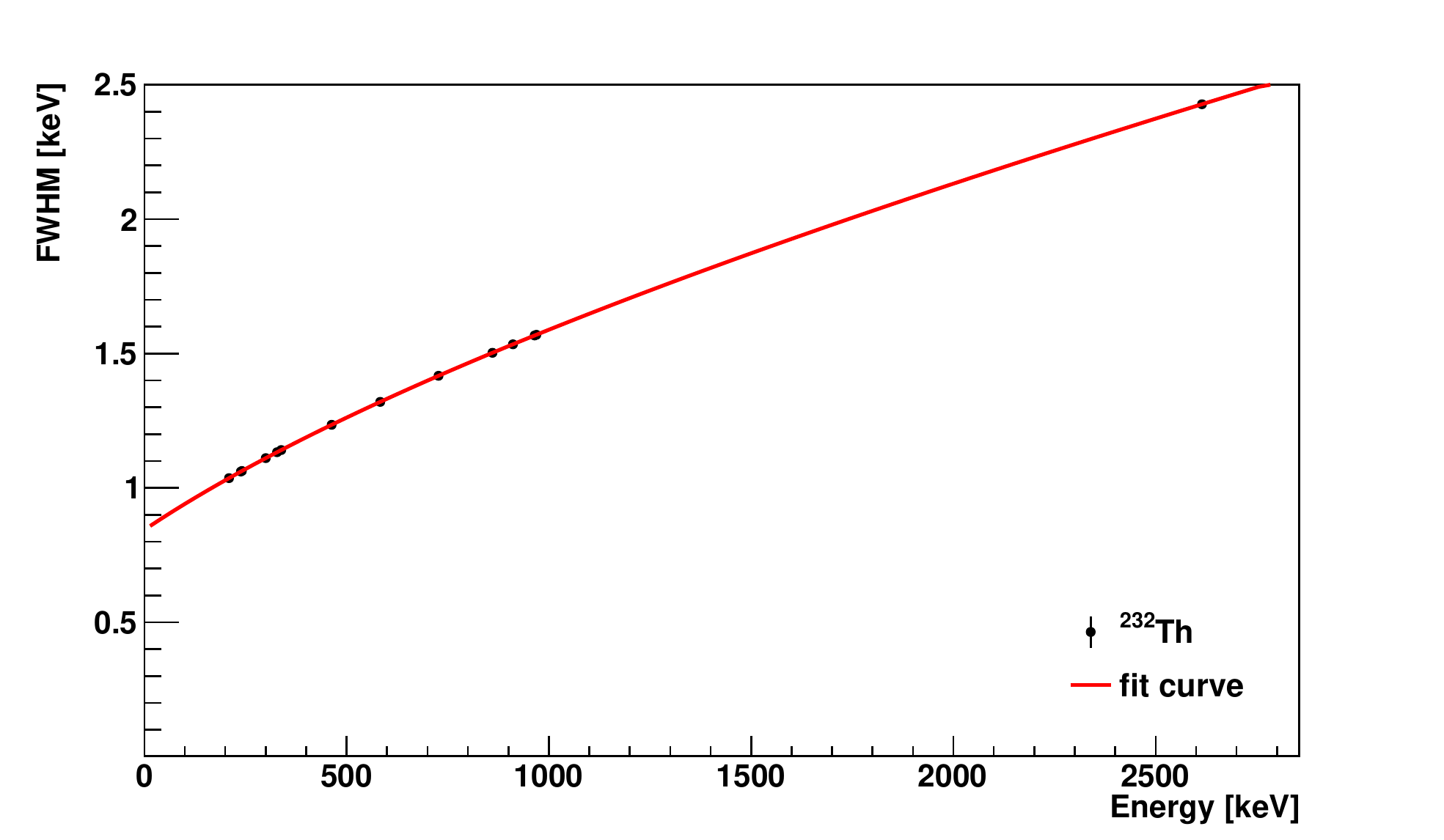}
\caption[]{Energy resolution of the detector system (defined here as FWHM) as a function of the incident gamma ray energy. The uncertainties of the data points are barely visible. The solid curve represents a fit using the parametrization of \autoref{eq:resol}.}
\label{fig:resol}
\end{figure}

The energy resolution of the germanium detector depends on the incident photon energy and is here defined as full width at half maximum of the peak (FWHM) in \si{\keV}. 
In \autoref{fig:resol} the FWHM $\omega$ of the Gaussian fit functions for each peak of the \isotope[232]{Th} calibration source is shown together with a curve parameterized as
\begin{equation}
\omega^2\left(E\right) = e^2 + p^2 E + c^2 E^2 ,
\label{eq:resol}
\end{equation}
where the parameters represent the electronic noise ($e$), the statistical fluctuation of the charge production ($p$), and the charge collection parameter ($c$) \cite{Gilmore.2008.}. 
The fit parameters describing the curve in \autoref{fig:resol} are: 
\begin{align*}
e&=\SI{0.8442\pm0.0016}{keV^2} \\
p&=\SI{0.0413\pm0.0001}{keV^{1/2}} \\
c&=\SI{0.000325\pm0.000007}{}
\end{align*}
The energy resolution of the DLB at \SI{1332.5}{keV} (\isotope[60]{Co}) is measured to be \SI[separate-uncertainty]{1.772(001)}{keV}, corresponding to a relative energy resolution of \SI{0.13}{\percent}. 

\subsection{Detection efficiency and simulation}
\label{sec:sim}

In order to determine the specific activity of a certain nuclide, the (absolute) full-energy peak detection efficiency $\epsilon$ of the system needs to be known precisely. 
It is strongly affected by the sample placement and geometry as well as self-absorption effects inside the sample and the surroundings.
The DLB uses MC simulations to calculate the FEP detection efficiency for each sample geometry individually. 
By this method, extensive sample preparation and measurements of (efficiency) calibration sources for each new sample configuration can be avoided.

The simulation program used for the DLB is VENOM, which has been developed by the COBRA collaboration and is based on the Geant4 Monte Carlo framework \cite{Agostinelli.2003.250}. 
VENOM makes use of the low-energy extensions of Geant4 and provides interfaces to generate particles or decays at a certain location within the implemented geometry. 
Furthermore, the relevant data produced during the tracking of particles by Geant4 is extracted and stored into structured ROOT \cite{Brun.1997.81} files for analysis.
The complete inner shielding, the surroundings of the detector, as well as some standard sample and source geometries are implemented in the virtual model of the detector system.  
Whereas the dimensions of the shielding are known, several parameters concerning the germanium detector, that are given by the manufacturer, need to be adapted. \\
Comparative studies were performed to match the simulated data with calibration measurements. 
The detector model is optimized in various characteristics of the germanium crystal, e.g. the dead layer thickness and the active crystal volume. 
Details on this procedure and the implementation can be found in Ref. \cite{neddermann_2014_phd}.

Since the mechanisms in VENOM do not contain a charge transport or electric field simulation, the MC simulation data is convoluted with the energy-dependent detector resolution, which is obtained from calibration measurements and using the parametrization defined in \autoref{eq:resol}. 

\begin{figure}[h!]
\includegraphics[width=1.08\columnwidth, angle=0]{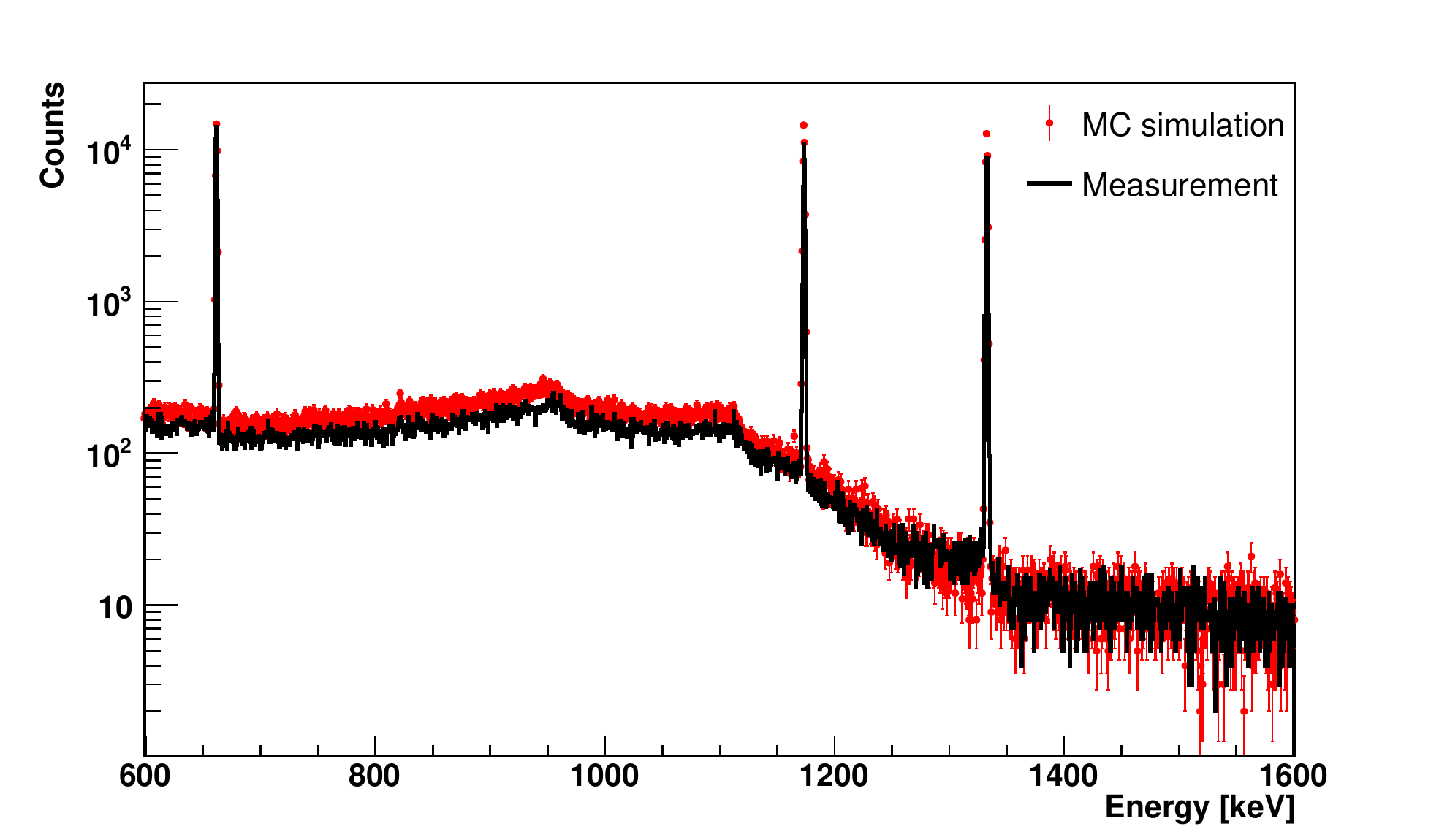}
\caption[]{Comparison of Monte Carlo simulation (red points) and measured data (black line) of a sample containing \isotope[60]{Co} and \isotope[137]{Cs}. Spectra shown in absolute counts.}
\label{fig:meas_vs_sim}
\end{figure}

\autoref{fig:meas_vs_sim} shows a comparison of a simulated spectrum and a measurement with the DLB. 
The spectrum is from a sample containing the isotopes \isotope[60]{Co} and \isotope[137]{Cs}.
Since the measured activity is relatively high in comparison to the background of the DLB, no background simulation is needed. 
Except for minor differences, no systematic deviations between the simulated and the measured data are observed.

\section{Background measurements and detection limits}
\label{sec:background}

In this section, the obtained background level of the DLB is discussed and the achieved detection limits are shown. 
The remaining background peaks are identified and analysed. 
Also the detection limits for several isotopes are calculated. 
The background count rate is discussed in the context of other European (ultra-)low-level gamma ray spectrometry laboratories. 

\subsection{Background-rate measurements}

\autoref{tab:bg_rates} summarizes several integral count rates measured at different stages of the construction of the DLB, and therefore gives an overview of the background reduction due to the different shielding installations, discussed in \autoref{sec:setup}. 
For each stage, the integral count rate is measured in the energy interval from \SI{40}{\keV} to \SI{2700}{\keV} and normalized to the germanium detector's mass.
This value is also referred to as the background index. 

The first value was measured inside the laboratory of the institute without any shielding arrangements. 
The second value was obtained with the outer shielding installed, but only with a temporary lead castle of \SI{5}{\cm} thickness surrounding the detector. 
The third count rate was measured with the current setup in terms of the passive outer and inner shielding, but without the active muon veto. 
The last value is the currently reached background index with passive shielding and muon veto.

\begin{table}[h!]
\centering
\caption[]{Integral background count rates between \SI{40}{\keV} and \SI{2700}{\keV} at different stages of construction of the DLB. The integrals are normalised to the germanium crystal mass.}
\label{tab:bg_rates}
\begin{tabular}{lS}
\toprule
\multicolumn{1}{l}{Stage} 		& \multicolumn{1}{c}{Integral count rate } \\
 							& \multicolumn{1}{c}{from \SIrange{40}{2700}{\keV} }\\
 							& \multicolumn{1}{c}{[\si{\counts\per\kg\per\minute}]}  \\
\midrule 
Unshielded					&	7681.4\pm1.3	\\ 
Outer shielding					&	101.95\pm0.18	\\
DLB without veto				&	23.888\pm0.040	\\ 
DLB with veto					&	2.5284\pm0.0035	\\ 
\bottomrule 
\end{tabular}
\end{table}

Compared to the un-shielded case, the current setup reduces the background by a factor of about \SI{3000}{}. 
The reduction due to the muon veto is a factor of \SI{9.5}{} at the current stage of development. 
After completing the veto-detector upgrade, a background index well below \SI{2}{\counts\per\kilo\gram\per\minute} is expected. 

\subsection{Background energy spectra}

\begin{figure}[h!]
\includegraphics[width=0.99\columnwidth, angle=0]{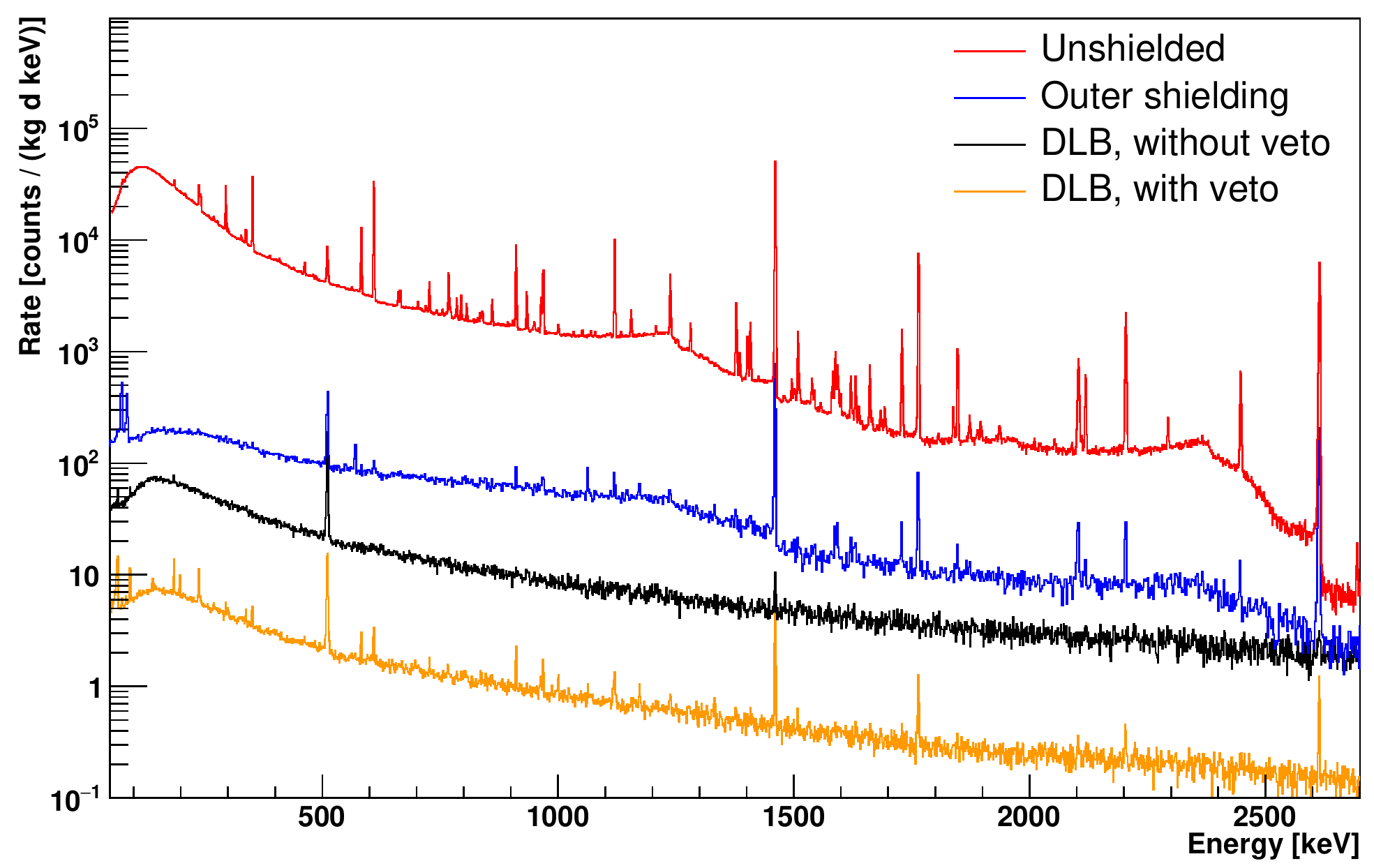}
\caption[]{Background spectra for \SI{40}{\keV} to \SI{2700}{\keV} at different stages during the construction of the DLB. Spectrum under unshielded conditions (red line), with the overburden of \SI{10}{\mwe} but without the inner shielding (blue line), completed passive shielding (black line) and current passive shielding with active muon veto (orange line).}
\label{fig:backgrounds}
\end{figure}

The different energy spectra shown in \autoref{fig:backgrounds} correspond to the integral count rates given in \autoref{tab:bg_rates}.

The spectrum that was obtained under unshielded conditions features a large number of gamma ray peaks originating from environmental background radiation. 
A dominant peak originating from \isotope[40]{K} is clearly visible as well as several gamma lines induced by the decay chains of \isotope[232]{Th} and \isotope[238]{U}.

The usage of the outer shielding in combination with a temporarily installed  lead shielding reduces the gamma background from the environment by nearly two orders of magnitude. 
Peaks from the uranium and thorium decay chain are strongly suppressed, but especially the thorium decay chain remains visible. 

After installation of the complete inner shielding, most gamma ray peaks are not visible in the fluctuations of the measured background spectrum.
Only the $e^+ e^-$ annihilation peak at \SI{511}{keV}, that is induced by cosmic muons, as well as the \isotope[40]{K} peak are visible.
Compared to the unshielded setup the count rate is lowered by a factor of approximately \SI{300}{}.

With applying the active muon veto more peaks become visible due to the remaining contamination as well as activation of the shielding and detector material. 
The overall count rate is lowered by nearly one order of magnitude.
The count rate drops below \SI{1}{\counts\per\kilo\gram\per\day\per\keV} above approximately \SI{1000}{\keV}.
Since the annihilation peak is mainly induced by cosmic muons hitting the vicinity of the germanium detector, the impact of the muon veto can be seen in the reduction of the net count rate in the peak area at \SI{511}{\keV}. 
The count rate is reduced from \SI{699\pm11}{\counts\per\day} with only the passive shielding in place to \SI{47.6\pm1.0}{\counts\per\day} with active muon veto activated. 

\autoref{tab:peaks} gives a summary of several peaks in the lowest-level background spectrum of the DLB, that was acquired for 110 days of measurement time during 2014 and 2015.
For the peaks that are not visible in the measured spectrum only upper limits for the count rate are given.
The net count rates of the full energy peaks are given in counts per day. 

\begin{table}[h!]
\centering
\caption[]{Integral count rate for several gamma ray energies of common nuclides. Integrals calculated from the background spectrum in counts per day. 'BE' denotes binding energy \cite{Isus_Table}. Upper limits with \SI{95}{\percent} C.L..}
\label{tab:peaks}
\begin{tabular}{SSc}
\toprule
{Energy [\si{\keV}]}	& {Peak [\si{\counts\per\day}]}  		& Nuclide / Chain 					\\
\midrule
46.54			& \num{9.89\pm0.52}			& \isotope[210]{Pb} (\isotope[238]{U})	\\
53.44			& \num{1.84\pm0.35}			& \isotope[73m]{Ge}					\\
66.72			& \num{16.54\pm0.87}			& \isotope[73m]{Ge}					\\
122.06			& <1.00 						& \isotope[57]{Co}					\\
136.47			& <0.54 						& \isotope[57]{Co}					\\
139.68			& \num{4.99\pm0.47}			& \isotope[75m]{Ge}					\\
143.57			& \num{2.66\pm0.51}			& \isotope[57]{Co} + \isotope{Fe} BE		\\
174.95			& <0.88 						& \isotope[71m]{Ge}					\\
198.4			& \num{6.46\pm0.45}			& \isotope[71m]{Ge}					\\
238.63			& \num{12.04\pm0.54}			& \isotope[212]{Pb} (\isotope[238]{U})	\\
295.22			& \num{2.59\pm0.36 }			& \isotope[214]{Pb} (\isotope[238]{U})	\\
351.93			& \num{4.77\pm0.42}			& \isotope[214]{Pb} (\isotope[238]{U})	\\
511				& \num{47.44\pm0.77}			& $e^+ e^-$ annih. 					\\
583.19			& \num{3.20\pm0.32}			& \isotope[208]{Tl} (\isotope[232]{Th})	\\
609.31			& \num{4.82\pm0.35}			& \isotope[214]{Bi} (\isotope[238]{U})	\\
661.6			& <0.51 						& \isotope[137]{Cs}					\\
691.4			& <0.27 						& \isotope[72]{Ge}					\\
810.77			& <0.24 						& \isotope[58]{Co}					\\
817.87			& <0.63 						& \isotope[58]{Co} + \isotope{Fe} BE		\\
834.85			& \num{0.62\pm0.20}			& \isotope[54]{Mn}					\\
840.84			& <0.29						& \isotope[54]{Mn} + \isotope{Cr} BE		\\
911.20			& \num{3.86\pm0.29}			& \isotope[228]{Ac} (\isotope[232]{Th})	\\
964.79			& <0.65 						& \isotope[228]{Ac} (\isotope[232]{Th})	\\ 
968.96			& \num{1.87\pm0.23}			& \isotope[228]{Ac} (\isotope[232]{Th})	\\
1001.03			& \num{1.35\pm0.20}			& \isotope[234m]{Pa} (\isotope[238]{U})	\\
1115.55			& <0.70 						& \isotope[65]{Zn}					\\
1120.29			& \num{1.69\pm0.27}			& \isotope[214]{Bi} (\isotope[238]{U})	\\
1124.53			& <0.21 						& \isotope[65]{Zn} + \isotope{Cu} BE		\\
1173.23			& \num{0.96\pm0.18}			& \isotope[60]{Co}					\\
1332.49			& \num{0.77\pm0.17}			& \isotope[60]{Co}					\\
1460.82			& \num{12.75\pm0.37}			& \isotope[40]{K}					\\
1764.49			& \num{2.81\pm0.21}			& \isotope[214]{Bi} (\isotope[238]{U})	\\
2204.21			& \num{0.88\pm0.16}			& \isotope[214]{Bi} (\isotope[238]{U})	\\
2614.51			& \num{4.06\pm0.24}			& \isotope[208]{Tl} (\isotope[232]{Th})	\\
\bottomrule
\end{tabular}
\end{table}

Several of the remaining peaks in the background spectrum are caused by capture of (thermal) neutrons, e.g. \isotope[72]{Ge}$\left(n,\gamma\right)$\isotope[73m]{Ge}, or inelastic scattering, e.g. \isotope[65]{Cu}$\left(n,n'\gamma\right)$\isotope[65]{Cu}, due to tertiary neutrons. 
Although the neutron flux is strongly suppressed by the passive shielding, it can not be stopped completely.  
For rather short living radionuclides, the corresponding gamma ray energy can be rejected from the spectrum by applying a rejection time of \SI{50}{\micro\second} after each muon veto event. 
For example, the \SI{174.9}{keV} line of \isotope[71m]{Ge}, with a half life of \SI{79}{\nano\second}, is suppressed effectively by using the muon veto of the DLB. 
Peaks originating from de-excitation with a half-life longer than the applied rejection time can not be reduced by this active reduction technique.
The height of these peaks, however, is influenced by the effectiveness of the neutron shielding. 

The de-excitation of the \SI{66.7}{keV} metastable state of \isotope[73]{Ge} does occur by the sequential emission of two gamma rays with \SI{53.4}{keV} and \SI{13.3}{keV}. 
Since the \SI{13.3}{keV} level has a half life of \SI{2.95}{\micro\second} \cite{Firestone.1996.} it is likely that the second transition occurs during the used shaping time of the main amplifier of \SI{6}{\micro\second}. 
Therefore, the peak at \SI{66.7}{keV}, representing the sum of both gamma rays, is much higher than the \SI{53.4}{keV} line and is the by far most intense line below \SI{500}{keV}. 

The flux of thermal neutrons within the detector shielding can be calculated using 
\begin{equation}
	\Phi_\mathrm{th} \left(\frac{\textrm{n}}{\si{\centi\metre\squared\second}}\right) = 
	\frac{R_{\mathrm{n},\,\SI{139.7}{\keV}}}{N(\isotope[74]{Ge})\, \sigma(\isotope[75]{Ge})\, 
	\frac{\varepsilon_{\SI{139.7}{\keV}} + 
	\alpha_\mathrm{tot}}{1 + \alpha_\mathrm{tot}}}\quad,
	\label{eq:ds:neutron_flux_thermal}
\end{equation}
where $R_{\mathrm{n}, \SI{139.7}{keV}}$ is the net count rate of the \SI{139.7}{keV} peak per second, which is induced by the \isotope[74]{Ge}$\left(n,\gamma\right)$\isotope[75m]{Ge} reaction \cite{Skoro.1992.333}.
$N(\isotope[74]{Ge})$ is the number of \isotope[74]{Ge} isotopes in the detector, which can be calculated from the detector mass and the natural abundance of \isotope[74]{Ge} of \SI{36.73(8)}{\percent} \cite{Boehlke.2005.57}. 
$\varepsilon_{\SI{139.7}{keV}}$ is the FEP efficiency for the \SI{139.7}{keV} gamma ray, which was determined by MC simulations, to be $\epsilon= 0.779\pm 0.039$. 
The de-excitation of \isotope[75m]{Ge} can alternatively occur by internal conversion, where the energy of the nucleus is transferred to an electron of a lower shell, which is then emitted from the atom. 
This effect is taken into account by adding the total conversion coefficient $\alpha_\mathrm{tot}= 1.54$ from Ref. \cite{Farhan.1999.785}, accessed via Ref. \cite{NuDat2}. 
$\sigma$(\isotope[74]{Ge}) is the partial neutron capture cross section for the production of the metastable state of \isotope[75]{Ge}: $\sigma\left(\isotope[74]{Ge}\right)= \SI{0.17(3)}{\barn}$ from Ref. \cite{Mughabghab.1981.}. \\
Using these data, the thermal neutron flux at the location of the HPGe detector is estimated to be about \SI[per-mode=reciprocal]{9.8\pm 0.9E-5}{\per\centi\metre\squared\per\second} during the spectrum acquisition period. 
This value is quite low compared to the value of approximately \SI[per-mode=reciprocal]{3E-3}{\per\centi\metre\squared\per\second} for a conventional shielding without a neutron moderator given in Ref. \cite{Skoro.1992.333} and proves the effectiveness of the neutron shielding. 

The inelastic scatter reaction \isotope[72]{Ge}$\left(n,n'\gamma\right)$\isotope[72]{Ge} is inducing a peak at \SI{691.4}{keV}. 
Using the width of the peak at from \cite{Skoro.1992.333}, one can calculate an upper limit on the count rate of this peak in the spectrum of \SI[per-mode=reciprocal]{5.7e-5}{\per\second}. 
The flux of fast neutrons can be calculated from the count rate $R_{\mathrm{n}}$ of the \SI{691.4}{keV} peak with the empirical formula
\begin{equation}
\Phi_\mathrm{fast}\left(\frac{\textrm{n}}{\si{\centi\metre\squared\second}}\right) =k\frac{\ R_{\mathrm{n},\SI{691.4}{keV}}}{V}\quad,
\label{eq:neutron_flux_fast}
\end{equation}
where $V$ is the volume of the detector in \si{\centi\metre\cubed} and $k$ an empirical factor of \num[separate-uncertainty]{900(150)} \cite{Skoro.1992.333}.
Using \autoref{eq:neutron_flux_fast}, the detector volume of \SI{234}{\centi\metre\cubed} and the count rate given above, an upper limit on the flux of fast neutrons of \SI[per-mode=reciprocal]{2.2E-4}{\per\centi\metre\squared\per\second} can be determined. 
This calculated value is nearly two orders of magnitude smaller than the value of \SI[per-mode=reciprocal]{1.7E-2}{\per\centi\metre\squared\per\second} given in Ref. \cite{Skoro.1992.333} for a conventional system without integrated neutron shielding. 
It has to be noted that the calculated upper limit is five times lower then the value given in Ref. \cite{Niese2008209} for the Felsenkeller with an overburden of \SI{110}{\mwe}.
Furthermore, the limit is nearly equal to the upper limit of \SI[per-mode=reciprocal]{2}{\per\metre\squared\per\second} given by \cite{Wordel.1996.557} for a system with \SI{500}{\mwe} of overburden. 
This again demonstrates the background-reduction capability of the DLB with respect to neutrons. 

Several peaks in the background spectrum are associated to the aforementioned activation by cosmic ray particles. 
The activated nuclides can be sorted into two groups, those which are produced in germanium and those created in the surroundings of the detector, e.g. copper or steel. 
They can be separated by the fact that the decay of the intrinsic nuclide is visible in a peak with higher energy, since the binding energy (BE) of the daughter nuclide is also deposited in the detector crystal \cite{neddermann_2014_phd}.
Four of the aforementioned activation products in germanium are \isotope[57]{Co}, \isotope[58]{Co}, \isotope[54]{Mn} and \isotope[65]{Zn}, having half lives in the range from \SI{70}{\day} to \SI{313}{\day}. 
The intrinsic decays of these nuclides can be seen as peaks in the spectrum at \SI{143.6}{\keV}, \SI{817.9}{\keV}, \SI{840.8}{\keV} and \SI{1124.5}{\keV}, respectively. 
After determining the full energy detection efficiency via MC simulation, the specific activities are calculated and shown in \autoref{tab:ge_activation}.
Ref. \cite{neddermann_2014_phd} gives a detailed analysis of the background spectrum that was acquired in 2009, of which the results of the analysis of the activation products in germanium are shown in \autoref{tab:ge_activation} a well.
As it was stated, especially the activities of \isotope[57]{Co} and \isotope[65]{Zn} are much smaller in 2014/15, due to their half lives of \SI{271.8}{\day} and \SI{244.3}{\day}.
Since the detector was permanently surrounded by its shielding containing an integrated neutron moderator, the flux of fast neutrons created by muon interactions in shielding materials has been noticeably reduced over the last five years of operation. 

\begin{table}
\caption[]{	Activities of activation products in the germanium detector derived from the background spectra acquired end of 2009 \cite{neddermann_2014_phd} and 2014/15. Efficiencies determined by MC simulations. Upper limits are given at \SI{95}{\percent} C.L..}
\label{tab:ge_activation}
\centering
\begin{tabular}{lSS}
\toprule
\multicolumn{1}{l}{Nuclide}	&\multicolumn{1}{c}{Specific activity}	&\multicolumn{1}{c}{Specific activity}		\\
						&\multicolumn{1}{c}{measured in 2009}	&\multicolumn{1}{c}{measured in 2014/15}	\\
						& [\si{\micro\becquerel\per\kilogram}]	& [\si{\micro\becquerel\per\kilogram}]		\\
\midrule
\isotope[57]{Co}			& \num{77\pm16}					& \num{34\pm19}						\\
\isotope[58]{Co}			& <45							& <44								\\
\isotope[54]{Mn}			& <49							& <17								\\
\isotope[65]{Zn}			& \num{332\pm91}					& <33								\\
\bottomrule
\end{tabular}
\end{table}

This reduction of activation due to cosmic ray particles can be seen in several other peaks as well. 
\isotope[57]{Co} and \isotope[58]{Co} are produced in copper as well and are visible as peaks at energies of \SI{122.1}{\keV}, \SI{136.5}{\keV} and \SI{810.8}{\keV} (\isotope[58]{Co}).
Additionally, \isotope[60]{Co} is produced in copper and visible as peaks at \SI{1173.2}{\keV} and \SI{1332.5}{\keV}. 
Since the detector has a cooling finger made of stainless steel, this detector part could be responsible for the \isotope[60]{Co} seen in the spectrum as well.

The remaining gamma peaks shown in \autoref{tab:peaks} are induced by the primordial nuclides and long-lived decay chains. 
In comparison to the background spectrum acquired in 2009, no reduction in the count rate can be observed, since these contaminations are not caused by activation processes and therefore not influenced by the upgrade of the muon veto or the operation time under shielded conditions.

\subsection{Detection Limits}

\autoref{tab:limits} gives the detection limits of the DLB according to \mbox{DIN ISO 11929:2011} \cite{DINISO.11929}, which is the German version of the international standard \mbox{ISO 11929:2010} and considers the uncertainties of the measurement as well as the one from weighing, sample treatment, efficiency calculations etc. (see Ref. \cite{neddermann_2014_phd} for details).  
The calculations are done for a typical sample volume ($\emptyset =$\SI{40}{\milli\metre}, height = \SI{40}{\milli\metre}) consisting of water. 
All results are calculated for a standard counting time of seven days.
By increasing the measurement time, the sensitivity is expected to scale with $\sim t^{1/2}$. 
After, for example, 28 days of measurement the detection limit is as low as a few \si{\milli\becquerel\per\kilo\gram} for some isotopes, e.g. \isotope[60]{Co} or \isotope[137]{Cs}.

\begin{table}[h!]
\centering	
\caption[]{Detection limits of the DLB for several isotopes. Calculations done for a typical sample volume ($\emptyset =$\SI{40}{\milli\metre}, height $=$ \SI{40}{\milli\metre}) of water and a counting time of seven days.}
\label{tab:limits}
\begin{tabular}{llSS}
\toprule 
\multicolumn{1}{c}{Chain}			& \multicolumn{1}{c}{Isotope} 	& \multicolumn{1}{c}{ Energy} 	& \multicolumn{1}{c}{ Detection limit}	\\
 							&						& [\si{keV}]				& [\si{\milli\becquerel\per\kilo\gram}]		\\
\midrule 
\isotope[238]{U}/\isotope[226]{Ra}	& \isotope[226]{Ra}			&	186.21				&	384.06						\\ 
							& \isotope[214]{Pb}			&	295.22				&	74.79						\\ 	
							& \isotope[214]{Pb}			&	351.93				&	42.37						\\ 	
							& \isotope[214]{Bi}			&	609.31				&	40.60						\\ 
							& \isotope[214]{Bi}			&	1120.29				&	139.08						\\ 
							& \isotope[214]{Bi}			&	1764.49				&	166.23						\\ 	
\midrule
\isotope[232]{Th}/\isotope[228]{Ra}	& \isotope[212]{Pb}			&	238.63				&	33.83						\\ 	
							& \isotope[208]{Tl}			&	583.19				&	19.93						\\ 	
							& \isotope[228]{Ac}			&	911.20				&	81.48						\\ 
							& \isotope[228]{Ac}			&	968.97				&	126.24						\\ 	
							& \isotope[208]{Tl}			&	2614.15				&	34.88						\\ 	
\midrule
							& \isotope[137]{Cs}			&	661.66				&	18.27						\\ 	
							& \isotope[60]{Co}			&	1173.23				&	19.55						\\ 	
							& \isotope[60]{Co}			&	1332.49				&	19.73						\\ 	
							& \isotope[40]{K} 			&	1460.83				&	344.62						\\ 	

\bottomrule 
\end{tabular}
\end{table}

The detection limit also scales with the square root of the background index. 
Hence, it is necessary to lower the background in order to achieve more sensitivity in a reasonable amount of measurement time. 

\subsubsection{Comparison with other laboratories}

The obtained results can be compared to other existing low-level gamma ray spectrometry laboratories in Europe. 
The background index from \SI{40}{\keV} to \SI{2700}{\keV} is normalized to the germanium detector's mass and given in \autoref{tab:comp} for several laboratories with various overburdens, ranging from \SI{1}{\mwe} to \SI{4800}{\mwe} of coverage. 

\begin{table*}
\begin{minipage}[c]{0.99\textwidth}
\centering
\caption[]{Comparison of the integral background count rate in the range of \SI{40}{\keV} to \SI{2700}{\keV} for several gamma ray spectrometry laboratories in Europe. The count rate is normalized to the germanium crystal mass and minute.}
\label{tab:comp}
\begin{tabular}{cccS}
\toprule
\multicolumn{1}{c}{Underground Laboratory} 	& \multicolumn{1}{c}{Institute} 							& \multicolumn{1}{c}{Depth} 	& \multicolumn{1}{c}{Integral count rate}			\\
 									&												&						& \multicolumn{1}{c}{ \SIrange{40}{2700}{\keV} }	\\
 									& 												&  [\si{\mwe}]				& [\si{\counts\per\kg\per\minute}]  				\\
\midrule
ARC 								& IAEA (Seibersdorf, Austria) \cite{Laubenstein.2004.167}	& 1						& \num{5.69\pm0.14} 		\\ 
DLB									& TU Dortmund (Dortmund, Germany)					& 10						& \num{2.5284\pm0.0035}	\\ 
LLL									& MPIK (Heidelberg, Germany) \cite{Heusser:2015ifa}		& 15						& \num{0.242\pm0.002}  		\\ 
CAVE								& IAEA-MEL (Monaco, Monaco) \cite{Povinec2006538}		& 35						& \num{0.58} 				\\ 
Felsenkeller							& VKTA (Dresden, Germany) \cite{Koehler2009736}		& 110					& \num{2.04} 				\\ 
HADES								& IRMM (Mol, Belgium) \cite{Koehler2009736}				& 500					& \num{0.168} 				\\ 
UDO									& PTB (Braunschweig, Germany) \cite{Neumaier2009726}	& 2100					& \num{0.31} 				\\ 
LNGS 								& LNGS (Assergi, Italy) \cite{Heusser:2015ifa}				& 3800					& \num{0.0458\pm0.0007} 	\\ 
LSCE 								& LSCE (Mondane, France) \cite{Laubenstein.2004.167}		& 4800					& \num{0.1292\pm0.0001}	\\ 
\bottomrule
\end{tabular}
\end{minipage}
\end{table*}

The achieved background level is comparable to that of the Verein für Kernverfahrenstechnik und Analytik (VKTA) laboratory Felsenkeller at Dresden-Rossendorf, Germany. 
This laboratory is covered by an overburden of \SI{110}{\mwe}, about one order of magnitude more than the DLB. 
However, no active muon veto is used in this facility \cite{Koehler2009736}.
The background index obtained by the Austrian Research Center (ARC) in Seibersdorf, Austria, is more than twice as high as the value achieved with the DLB \cite{Laubenstein.2004.167}.
Although the ARC is covered by less overburden, a muon veto is used as well. 
A lower background index has been achieved in the Low-Level Laboratory (LLL) in Heidelberg, Germany, where \SI{15}{\mwe} overburden are covering a large inner shielding, that was built using extensive studies for material selection for shielding and detector housing \cite{Heusser:2015ifa}.

\section{Summary}
\label{sec:summary}

The described low-level germanium spectrometer DLB is sensitive in the range of some \SI{10}{\milli\becquerel\per\kilo\gram}, depending on the measurement time.
It is therefore 20 to 50 times more sensitive than a gamma ray spectrometry laboratory without special shielding arrangements.   
The integral background count rate normalized to the germanium crystal mass in the energy range from \SI{40}{\keV} to \SI{2700}{\keV} is determined to be \SI{2.5284\pm0.0035}{\counts\per\kilo\gram\per\minute}, which is comparable to laboratories with a higher overburden.

The DLB is an ideal tool for low-level gamma ray spectrometry, e.g. material screening, radio assaying or environmental analytics. 
Due the operation of the DLB above ground and the low amount of overburden, access times for sample replacement is small. 
It thus combines the advantages of a standard gamma ray spectrometer in a conventional laboratory and those of low-background laboratories.  

\section*{Acknowledgement}

The authors thank the Deutsche Forschungsgemeinschaft (DFG) for supporting the experiment (grant number: GO 1133/1-2). 
The Alpha Guard detector for radon measurements inside the DLB was kindly provided by the Zentrales Isotopenlabor (ZIL), RUBION, Universität Bochum.

\section*{References}

\bibliography{DLB_paper}

\begin{thebibliography}{10}
\expandafter\ifx\csname url\endcsname\relax
  \def\url#1{\texttt{#1}}\fi
\expandafter\ifx\csname urlprefix\endcsname\relax\def\urlprefix{URL }\fi
\expandafter\ifx\csname href\endcsname\relax
  \def\href#1#2{#2} \def\path#1{#1}\fi

\bibitem{Cline:2014fua}
D.~Cline, {A brief status of the direct search for {WIMP} dark matter }\href
  {http://arxiv.org/abs/1406.5200} {\path{arXiv:1406.5200}}.

\bibitem{Barabash:2010ie}
A.~S. Barabash, {Precise half-life values for two neutrino double beta decay},
  Phys. Rev. C81 (2010) 035501.

\bibitem{COBRA.2015.2}
J.~Ebert, et~al., {Results of a search for neutrinoless double-beta decay using
  the COBRA demonstrator }\href {http://arxiv.org/abs/1509.04113}
  {\path{arXiv:1509.04113}}.

\bibitem{COBRA.2015.1}
J.~Ebert, et~al., {The COBRA demonstrator at the {LNGS} underground
  laboratory}, Nucl. Instrum. Meth. A807 (2016) 114--120.

\bibitem{borexino}
G.~Bellini, et~al., {Cosmic-muon flux and annual modulation in Borexino at 3800
  m water-equivalent depth}, JCAP 1205 (2012) 15.

\bibitem{Heusser.2006.495}
G.~Heusser, M.~Laubenstein, H.~Neder, {L}ow-level germanium gamma-ray
  spectrometry at the $\mu${B}q/kg level and future developments towards higher
  sensitivity, in: P.~Povinec, J.~Sanchez-Cabeza (Eds.), Radionuclides in the
  Environment - Int. Conf. On Isotopes in Env. Studies, Vol.~8, Elsevier, 2006,
  pp. 495 -- 510.

\bibitem{CRC.2015}
W.~Haynes, CRC Handbook of Chemistry and Physics, 96th Edition, CRC Press,
  2015.

\bibitem{Heusser.1995.543}
G.~Heusser, {Low-radioactivity background techniques}, Ann. Rev. Nucl. Part.
  Sci. 45 (1995) 543--590.

\bibitem{Firestone.1996.}
R.~B. Firestone, {T}able of {I}sotopes, 8th Edition, Vol. I \& II, John Wiley
  \& Sons, 1996.

\bibitem{Gilmore.2008.}
G.~R. Gilmore, {P}ractical gamma-ray spectrometry, 2nd Edition, John Wiley \&
  Sons, Ltd, 2008.

\bibitem{Koehler.2004}
M.~Koehler, et~al., Reference measurements of low levels of \isotope[60]{Co} in
  steel, Appl. Radiat. Isot. 61 (2004) 207 -- 211.

\bibitem{Povinec.2008}
P.~Povinec, et~al., New isotope technologies in environmental physics, Acta
  Physica Slovaca 58 (2008) 1 -- 154.

\bibitem{Niese2008209}
S.~Niese, Underground laboratories for low-level radioactivity measurements,
  in: Analysis of Environmental Radionuclides, Vol.~11, Elsevier, 2008, pp. 209
  -- 239.

\bibitem{Paul.1971}
J.~M. Paul, {T}he density effect and rate of energy loss in common plastic
  scintillators, Nucl. Instrum. Meth. A96 (1971) 51--59.

\bibitem{Brodzinski1990337}
R.~L. Brodzinski, et~al., {Further reduction of radioactive backgrounds in
  ultrasensitive germanium spectrometers}, Nucl. Instrum. Meth. A292 (1990)
  337--342.

\bibitem{Laubenstein2009750}
M.~Laubenstein, G.~Heusser, Cosmogenic radionuclides in metals as indicator for
  sea level exposure history, Appl. Radiat. Isot. 67 (2009) 750 -- 754.

\bibitem{Laubenstein.2004.167}
M.~Laubenstein, et~al., {U}nderground measurements of radioactivity, Appl.
  Radiat. Isot. 61 (2004) 167 -- 172.

\bibitem{CanberraIndustries.2009.SEGe}
{Canberra Industries, Inc.},
  \href{http://www.canberra.com/products/detectors/pdf/SEGe-SS-C37419.pdf}{{S}tandard
  {E}lectrode {C}oaxial {G}e {D}etectors ({SEG}e)}, Website (2009).
\newline\urlprefix\url{http://www.canberra.com/products/detectors/pdf/SEGe-SS-C37419.pdf}

\bibitem{CanberraIndustries.2010.drawing}
{Canberra Industries, Inc.}, {I}nner drawing p-type, {S}/{N} 07064, Data sheet
  (2010).

\bibitem{neddermann_2014_phd}
T.~Neddermann, \href{http://hdl.handle.net/2003/33482}{{Material Screening by
  means of Low-level Gamma Ray Spectrometry with the Dortmund Low Background
  HPGe Facility}}, Ph.D. thesis, Technische Universität Dortmund (2014).
\newline\urlprefix\url{http://hdl.handle.net/2003/33482}

\bibitem{CanberraIndustries.2008.ULB}
{Canberra Industries, Inc.},
  \href{http://www.canberra.com/products/detectors/pdf/Ultra-LowBkgrnd-Cryo-SS-CSP0173.pdf}{{U}ltra
  {L}ow-{B}ackground {C}ryostats ({ULB})}, Website (2008).
\newline\urlprefix\url{http://www.canberra.com/products/detectors/pdf/Ultra-LowBkgrnd-Cryo-SS-CSP0173.pdf}

\bibitem{Theodorsson.1996.}
P.~Theod\'{o}rsson, {M}easurement of weak radioactivity, World Scientific,
  1996.

\bibitem{Isus}
W.~Wahl, \href{http://www.isus.de/}{{I}nstitute for {S}pectrometry and
  {R}adiation {P}rotection}, Website (2008).
\newline\urlprefix\url{http://www.isus.de/}

\bibitem{Wieser.2009.2131}
M.~E. Wieser, M.~Berglund, {A}tomic weights of the elements 2007 ({IUPAC}
  {T}echnical {R}eport), Pure and Applied Chemistry 81 (2009) 2131--2156.

\bibitem{Crystals2009}
{Saint-Gobain Crystals}, Datasheet {BC}-408 (2009).

\bibitem{Photonics1992}
{Hamamatsu Photonics}, Datasheet photomultiplier tube {R}2490-05 (1992).

\bibitem{Agostinelli.2003.250}
S.~Agostinelli, et~al., {GEANT4: A Simulation toolkit}, Nucl. Instrum. Meth.
  A506 (2003) 250--303.

\bibitem{Brun.1997.81}
R.~Brun, F.~Rademakers, {ROOT} --- {A}n object oriented data analysis
  framework, Nucl. Instrum. Meth. A389 (1997) 81--86.

\bibitem{Isus_Table}
W.~Wahl, $\alpha\beta\gamma$-{T}able, Handbook (2007).

\bibitem{Skoro.1992.333}
G.~P. {\v{S}}koro, et~al., {E}nvironmental neutrons as seen by a germanium
  gamma-ray spectrometer, Nucl. Instrum. Meth. A316 (1992) 333 -- 336.

\bibitem{Boehlke.2005.57}
J.~K. Böhlke, et~al., {I}sotopic {C}ompositions of the {E}lements, J. Phys.
  Chem. Ref. Data 34 (2005) 57--67.

\bibitem{Farhan.1999.785}
A.~R. Farhan, B.~Singh, {N}uclear {D}ata {S}heets for {A} = 75, Nuclear Data
  Sheets 86 (1999) 785 -- 954.

\bibitem{NuDat2}
{Brookhaven National Laboratory},
  \href{http://www.nndc.bnl.gov/nudat2/}{{NuDat} 2}, Website.
\newline\urlprefix\url{http://www.nndc.bnl.gov/nudat2/}

\bibitem{Mughabghab.1981.}
S.~F. Mughabghab, M.~Divadeenam, N.~E. Holden, {N}eutron {C}ross {S}ections
  from {N}eutron {R}esonance {P}arameters and {T}hermal {C}ross {S}ections,
  Academic Press, 1981.

\bibitem{Wordel.1996.557}
R.~Wordel, et~al., {S}tudy of neutron and muon background in low-level
  germanium gamma-ray spectrometry, Nucl. Instrum. Meth. A369 (1996) 557--562.

\bibitem{DINISO.11929}
{B}estimmung der charakteristischen {G}renzen ({E}rkennungsgrenze,
  {N}achweisgrenze und {G}renzen des {V}ertrauensbereichs) bei {M}essungen
  ionisierender {S}trahlung -- {G}rundlagen und {A}nwendungen (2011).

\bibitem{Heusser:2015ifa}
G.~Heusser, et~al., {GIOVE - A new detector setup for high sensitivity
  germanium spectroscopy at shallow depth}, Eur. Phys. J. C75 (2015) 531.

\bibitem{Povinec2006538}
P.~Povinec, et~al., {IAEA}-{MEL}'s underground counting laboratory -- {T}he
  design and main characteristics, in: Radionuclides in the Environment - Int.
  Conf. On Isotopes in Env. Studies, Vol.~8, Elsevier, 2006, pp. 538 -- 553.

\bibitem{Koehler2009736}
M.~Koehler, et~al., A new low-level gamma-ray spectrometry system for
  environmental radioactivity at the underground laboratory {F}elsenkeller,
  Appl. Radiat. Isot. 67 (2009) 736 -- 740.

\bibitem{Neumaier2009726}
S.~Neumaier, et~al., Improvements of a low-level gamma-ray spectrometry system
  at the underground laboratory {UDO}, Appl. Radiat. Isot. 67 (2009) 726 --
  730.

\end{thebibliography}
\bibliographystyle{elsarticle-num}

\end{document}